\documentclass{article}

\usepackage[margin=1.25in]{geometry}
\usepackage[small,bf]{caption}
\usepackage{amsmath}
\usepackage{amssymb,url,graphicx,booktabs,epstopdf,appendix,textcomp,mathrsfs,epsfig,array}
\usepackage{graphicx}
\usepackage{amsfonts}
\usepackage{amsmath}
\usepackage{lineno}
\usepackage{placeins}
\usepackage{float}
\restylefloat{table}
\usepackage{verbatim}
\usepackage{multirow}
\usepackage[table]{xcolor}
\usepackage{tikz}
\usetikzlibrary{arrows}
\usepackage[english]{babel}
\usepackage{graphicx}
\usepackage{float}
\restylefloat{table}
\usepackage{verbatim}
\usepackage{multirow}
\usepackage{caption}
\usepackage{subfig}

\usepackage{lscape}

\usepackage{amssymb}

\usepackage{amsmath}
\usepackage{amssymb,url,graphicx,booktabs,epstopdf,textcomp,mathrsfs,epsfig,array,setspace}
\usepackage{graphicx}
\usepackage{amsfonts}
\usepackage{amsmath}
\usepackage{lineno}
\usepackage{placeins}
\usepackage{float}
\restylefloat{table}
\usepackage{verbatim}
\usepackage{multirow}
\usepackage{caption}
\usepackage{subfig}
\usepackage{amsmath}
\usepackage{tikz}
\usepackage{mathdots}
\usepackage{mathtools}
\usepackage{yhmath}
\usepackage{cancel}
\usepackage{color}
\usepackage{siunitx}
\usepackage{array}
\usepackage{multirow}
\usepackage{amssymb}
\usepackage{gensymb}
\usepackage{tabularx}
\usepackage{booktabs}
\usetikzlibrary{fadings}
\usepackage{natbib}
\usepackage{lineno}

\begin{document}

\title{Modeling the Role that Habitat Overlap Shape has on the Spread of Brucellosis in the Greater Yellowstone Ecosystem
}

\author{Dustin Padilla\medskip\\
\footnotesize{Simon A. Levin Mathematical, Computational Sciences Modeling Center}\\
 \footnotesize{Arizona State University, Tempe, AZ}}

\date{\today}

\maketitle

\begin{abstract}

Brucellosis is a zoonotic bacterial infectious disease that affects livestock and wildlife. The Greater Yellowstone Ecosystem is the last area in the United States where cattle are regularly infected with brucellosis. Even though livestock are vaccinated, interactions with reservoir species still result in spill-over cases to cattle. The National Academy of Sciences, Engineering, and Medicine has indicated that modeling efforts should focus on the transmission between elk and cattle and should be conducted to better understand the effects of land-use change and landscape configuration on disease risk. This chapter determines how the landscape's configuration, using the shape and amount of habitat overlap between elk and cattle, contributes to cross-species brucellosis transmission, and how land-use change translates to disease prevalence. A mathematical epidemiological model is combined with landscape ecology metrics to estimate transmission rates between the species and model disease spread. The results of this study can ultimately help stakeholders develop policy for controlling brucellosis transmission between livestock and elk in the Greater Yellowstone Area. In turn, this could lead to less disease prevalence, reduce associated costs, and assist in population management.

\end{abstract}

%\newpage

\section{Introduction}

Researchers have identified land-use change and the ensuing habitat fragmentation as central components in zoonotic and epizootic disease emergence, transmission, and persistence \citep{daszak2000emerging, patz2004unhealthy, wolfe2007origins}. However, the association of land-use change and landscape configuration with disease spread lacks mathematical theory  \citep{faust2018pathogen, suzan2008effect, suzan2012habitat}. In Chapter 2, a general theoretical framework about how the shape and size of a habitat fragment and the amount of habitat overlap on a landscape impact disease transmission and persistence was developed. The results provide insights into how the introduction of a disease free-population into an area, where an infectious disease is endemic to another population, will proliferate disease prevalence throughout both populations due to land conversion. However, the model was not parameterized for specific populations and a particular disease. In Chapter 3, an application of this framework is presented to demonstrate its ability to approximate disease prevalence as a result of land-use change, as well as further the understanding about the relationship between patterns of landscape configuration and disease spread.

For this chapter, the general framework is contextualized to study brucellosis transmission between elk and cattle in the Greater Yellowstone Ecosystem (GYE). Influenced by the National Academy of Sciences, Engineering, and Medicine (NASEM) report \textit{Revisiting Brucellosis in the Greater Yellowstone Area}, this case study strives to understand how landscape configuration and land-use change impact enzootic disease spread. \citet{national2017revisiting} is a comprehensive, peer-reviewed report that provides an evidence-based consensus about brucellosis in the GYE, and includes findings, conclusions, and recommendations based on information gathered and deliberations formed by a committee of experts. The report determined elk to be the primary source of new cases in cattle and recommended modeling transmission of brucellosis between the species, incorporating how land-use change and its effects on landscape configuration impact transmission risk \citep{national2017revisiting}.

%As a mechanistic model has not been presented to address these issues, the objective of this chapter is to model transmission between elk and cattle by using the model developed in the previous chapter to estimate how land-use change contributes to disease prevalence. Moreover, an estimation of transmission rates is conducted by determining the shape and size of the habitat overlap between cattle and elk in the GYE. Variation in that overlap translates to how land-use change impacts disease spread. The study adds to the existing research by responding to the NASEM report, providing brucellosis prevalence estimations for cattle and elk as a result of land-use change and the subsequent landscape configuration, and provides policy implications based on those results.

A mechanistic model to address these issues does not currently exist. This chapter adapts general the model developed in the previous chapter to estimate how land-use change contributes to disease prevalence in the GYE. It shows how transmission rates are affected by the shape and size of the habitat overlap between cattle and elk in the GYE, and therefore how land-use change impacts disease spread. The study adds to the existing research by responding to the \citet{national2017revisiting} report, providing brucellosis prevalence estimations for cattle and elk as a result of land-use change and the subsequent landscape configuration. It includes a discussion of policy implications based on those results.

\subsection{Background}

Brucellosis is a zoonotic bacterial infection caused by species of \textit{Brucella} \citep{corbel1997brucellosis, national2017revisiting}. The disease is found globally in numerous mammalian species, including humans \citep{franco2007human, young1995overview}. Brucellosis infection causes fever, gastrointestinal dysfunctionality, neurological disorders, and even death \citep{young1995overview}. Moreover, infection in females usually results in a reduction of fecundity, decreased milk production, and late-gestation abortion; about 70\% of infected pregnancies result in abortions \citep{franco2007human}. Transmission between individuals is highly contagious, and is primarily spread via contact with discharged contaminated matter or interaction with an infected animal \citep{rhyan2009pathogenesis}. Brucellosis can also be vertically transmitted, that is from mother to offspring during pregnancy \citep{corbel1997brucellosis}. The disease is internationally regulated in numerous species \citep{corbel2006brucellosis,richey1997brucella}.

In the United States, brucellosis is regulated in livestock because of trade and public health concerns \citep{ragan2002animal}. Through eradication efforts, via vaccination and quarantine, brucellosis has substantially been eliminated in cattle (\textit{Bos taurus}) across most of the U.S. \citep{ragan2002animal}. However, the disease, primarily caused by \textit{Brucella abortus}, continues to be persistently detected within cattle herds in the Greater Yellowstone Ecosystem (GYE), as interactions with elk (\textit{Cervus canadensis}) drives incidence \citep{national2017revisiting}. Testing indicates that about 20\% of cattle and 10--40\% of elk in the GYE are infected with brucellosis. Evidence suggests that brucellosis was introduced to the native ungulate populations through interactions with cattle \citep{meagher1994origin} and is now endemic to elk and bison (\textit{Bison bison}) in the region \citep{scurlock2010status}.

The Greater Yellowstone Ecosystem (GYE), is a 12--22-million-acre geographic region that spans from northwestern Wyoming, to northeastern Idaho, and southern Montana \citep{cross2010probable, aune2012environmental}. It is made up of lowland grasslands, alpine coniferous forest, and alpine tundra, with approximately a third of the GYE being low-elevation grasslands \citep{hansen2002ecological, cheville1998brucellosis}. The area is comprised of two national parks, Yellowstone National Park (YNP) and Grand Teton National Park (GTNP), both managed by the National Park Service (NPS) \citep{national2017revisiting}. The GYE also encompasses six full or partial national forests managed by the United States Forest Service (USFS), as well as Bureau of Land Management (BLM) holdings, private lands, and reserved-tribal lands  \citep{national2017revisiting}.

\newpage

The GYE supports a wealth of biodiversity, including the last remaining free-roaming wild herd of bison, located in the national parks, and the largest elk herds in the contiguous U.S.  \citep{national2017revisiting, schumaker2012brucellosis}.
Approximately 25,000 elk, making up nine major herds, inhabit both public and private lands \citep{national2017revisiting}. The lowland grasslands are preferential feeding spots for elk during the winter, as elk generally migrate from alpine regions in the summer to lowland grasslands in the winter \citep{cross2007effects}. Moreover, the majority of the biodiversity in the GYE is located on or near the lowland grasslands, with most of this land being privately owned \citep{cross2010probable,national2017revisiting}.

Private land holdings in the GYE are primarily located on lowland grasslands, and have been subject to land-use change within the past 40 years \citep{hansen2002ecological, national2017revisiting, hansen2018trends}. These areas are mainly used for cattle ranching, with approximately 450,000 cattle in the GYE, and about 85\% of the operations are open-range grazing calf-cow beef producers \citep{peck2010bovine,national2017revisiting}. This has created habitat overlap between cattle and elk. Additionally, significant portions of public lands in the area that are inhabited by elk are leased from the federal government for cattle production, which has also caused more habitat overlap between the species \citep{brennan2017shifting, proffitt2011elk}.

Land-use change in the region has created a common habitat for elk and cattle and has facilitated cross-species disease transmission \citep{national2017revisiting, cross2010probable}. Since cattle interactions with elk contribute to disease persistence, the GYE is the last area in the U.S. where brucellosis is consistently problematic to cattle \citep{national2017revisiting, scurlock2010status}. This has caused cattle ranchers and government agencies to invest in managing and researching the disease \citep{national2017revisiting, aune2012environmental}. However, current research has not presented a mechanistic model to \newpage \noindent describe how land-use change and the subsequent habitat overlap configuration facilitate cross-species disease transmission \citep{national2017revisiting}.

\subsection{Brucellosis Research in the GYE}

The goal of brucellosis research and management in the GYE is to ultimately eliminate the disease from the region \citep{national2017revisiting, kilpatrick2009wildlife}. However, eradicating the disease has been challenging, and multiple strategies exist on how to manage the disease in wild and domesticated species \citep{national2017revisiting, cross2007effects}. Cattle vaccination against brucellosis has been a vital component in reduction of disease incidence and is highly effective when administered in conjunction with other disease control efforts, such as quarantining infected individuals and managing herd mixing \citep{national2017revisiting}. Although vaccination is used to help prevent the transmission of brucellosis within domesticated populations and, as a result, provides partial protection to wild populations through herd immunity, it is not 100\% effective \citep{national2017revisiting, cross2007effects}. This lack of efficacy creates the necessity for other management strategies to reduce transmission--one of the most apparent being spatial-temporal separation \citep{national2017revisiting, rayl2019modeling}. 

Brucellosis incidence in cattle was initially suspected to occur primarily from interactions with bison, since the species can cross-breed and exhibit similar foraging behavior \citep{meagher1994origin}. Spatial-temporal separation of the species has successfully reduced brucellosis transmission from bison to cattle in the GYE, as bison have been restricted to the national parks \citep{national2017revisiting}. As a result, no further documented spill-over cases from bison have been recorded since the policy was implemented \citep{national2017revisiting}. 

The \citet{national2017revisiting} report concluded that elk are the primary source of new brucellosis cases to cattle in the Designated Surveillance Area (DSA) of the GYE, which was specified in the report. The report recommended that modeling should be used to characterize and quantify the risk of infection from elk to cattle, and suggested that studies incorporate how land-use change impacts spatial-temporal processes associated with disease spread. Moreover, it recommended that brucellosis management strategies focus on increasing the spatial-temporal separation of the species, thus limiting species interactions and potential cross-species disease transmission. As current studies have not accounted for these issues, the \citet{national2017revisiting} report's recommendations motivate this study.

Land-use change and its impacts on landscape configuration affect the spatial-temporal processes associated with the epidemiology of brucellosis  \citep{national2017revisiting,hansen2009species}. Land-use change has caused a transition in the amount of habitat overlap, thereby altering species' interactions through an adjustment in the types of species on the landscape, by a variation in population densities, or by an alteration in migration routes \citep{meentemeyer2012landscape,cross2007effects,cotterill2018winter}. All of these factors have been shown to impact disease transmission and complicate spatial-temporal separation of the species.

Land-use change impacts the types of species that inhabit a landscape. Expansion of land allocated for cattle grazing introduces cattle to areas that are inhabited by elk; this results in habitat overlap between these species, more species interactions, and ultimately in an increase of cross-species disease transmission \citep{brennan2014multi,hansen2009species}. \citet{cross2010probable} points out that while habitat overlap is critical to these interactions, the degree to which  the overlap contributes to cross-species brucellosis transmission is an open question. \citet{rayl2019modeling}, estimated that 98\% of cross-species brucellosis transmission occurs in lowland grasslands (the majority of which are private lands) due to habitat overlap caused by cattle ranching. 
\newpage

Transmission in lowland grasslands is significant because each species is exposed to brucellosis at a time of high seroprevalence when cross-species interactions are likely \citep{cross2007effects,national2017revisiting}. Cross-species and within-species transmission primarily occurs during the birthing season \citep{dobson1996population}. The birthing season often aligns with elks' migration to lowland grasslands, where they frequently interact with cattle \citep{cross2010mapping}. When the species interact with infected calves or aborted fetal remnants, the high density of \textit{Brucella} on the placenta has the ability to infect even vaccinated individuals, which is considered the dominant reason for increased transmission within a population \citep{proffitt2011elk}. Species interactions at these times of high seroprevalence are suspected to result in cross-species infections, especially when densely situated \citep{proffitt2011elk,cross2010mapping}.

Land-use change can also alter population densities of species on a landscape \citep{suzan2012habitat,greer2008habitat,saunders1991biological}. The implementation of feedgrounds for elk is an example of this process \citep{cotterill2018winter}. As the lowland grasslands are elks' seasonal preferential feeding spots, in an effort to deter elk from interacting with cattle and reduce the risk of brucellosis transmission while still maintaining population abundance, federal and state agencies have created supplementary feedgrounds for elk \citep{cotterill2018winter}. It has been determined that usage of the feedgrounds exacerbates brucellosis prevalence due to an increased density of elk. The increased population density happens either by decreasing the spatial distribution at which elk forage or by increasing the relative abundance of that population, or both. During the birthing season, a time of high transmission, elk congregate at the feedgrounds, decreasing their spatial distribution. This increased population density at a time of high disease transmission increases the likelihood of infection in elk, and in turn is speculated to result in higher transmission rates to cattle when interactions occur \citep{cotterill2018winter,rayl2019modeling}. Elks' relative abundance changes at the supplemental feedgrounds, during a time of high transmission, and it has been shown that a higher abundance of the species yields higher disease transmission rates \citep{cotterill2018winter,ferrari2002bison}. 

Land-use change, by way of cattle ranching or the subsequent elk feedgrounds, also affects migration patterns \citep{proffitt2011elk,national2017revisiting}. Species interactions are changed by altering migration patterns which, in turn, can impact disease transmission \citep{cross2010probable, manzione1998venezuelan}. Elk migration routes, in particular, have changed considerably over a thirty-year time period \citep{cross2007effects,parmenter2003land}. As more cattle ranching has been introduced to the area, elk have foraged in cattle rangeland and have interacted with cattle more frequently, resulting in more cross-species disease infection \citep{gosnell2006ranchland,brennan2017shifting}. Moreover, due to agricultural development, the population of top-level predators which would prey on elk has decreased to an extent that cattle and elk interact more consistently \citep{middleton2013animal,gosnell2006ranchland}. The lack of predation has caused overpopulation in elk, resulting in a change of their migration patterns and foraging ranges into cattle habitat \citep{nelson2012elk, white2013yellowstone}. To control elk abundance in the region, wolves were reintroduced, causing elk to further migrate onto cattle habitat in order to avoid predation \citep{fortin2005wolves,ripple2015trophic}. Even though the reintroduction of wolves decreased elk abundance, it also increased cross-species interactions and, as a result, cross-species pathogen transmission. Land-use change by means of elk feedgrounds also affects migration patterns and has a counter-intended effect on spill-over brucellosis cases to cattle \citep{cotterill2020disease}. The feedgrounds are utilized to deter elk from grazing where cattle are located \citep{cotterill2020disease,national2017revisiting}. This shifts elk's migrations patterns to feedgrounds; however, due to an increase in population density at a time of high transmission, there is a higher prevalence of brucellosis in elk, which, in turn, increases the likelihood of disease spill-over to cattle \citep{cross2010probable,cotterill2020disease}.

The state of the landscape plays a role in species' inhabitance, cross-species' interactions, and, as a consequence, the spread of their infectious diseases \citep{patz2004unhealthy,meentemeyer2012landscape}. Species interactions can be affected by either a change in the types of species on the landscape, a variation in population densities, or altered migration routes, all of which have been shown to impact disease transmission \citep{suzan2012habitat,cross2007effects,cotterill2018winter}. These spatial and temporal processes are connected to the epidemiology of the disease. As the \citet{national2017revisiting} report suggests, modeling efforts should take into consideration how land-use change and landscape configuration contribute to cattle and elk interacting more consistently.

Previous literature has mostly focused on modeling brucellosis transmission in bison and elk in the GYE \citep{dobson1996population,cotterill2020disease,abatih2015mathematical}. \citet{dobson1996population} modeled brucellosis transmission between bison and elk in YNP. The study incorporated vertical transmission in the model and estimated transmission rates among and between the species, but did not incorporate the effects that landscape configuration or land-use change have on the populations sharing the pathogen. Additionally, the study estimated brucellosis prevalence levels in each species but did not consider how the shape or size of the habitat overlap between the species factor into disease transmission. \citet{cotterill2020disease} modeled transmission in elk at the winter feedgrounds but did not examine transmission to cattle or how that land-use change drives disease prevalence in the species.  Based on\citet{dobson1996population}, transmission in bison was modeled by \citet{abatih2015mathematical}, and \citet{xie2009disease} modeled transmission between cattle and elk in Wyoming and included economic decisions and vaccination strategies. The past literature has not presented a mechanistic model to describe how land-use change and landscape configuration contribute to brucellosis transmission between elk and cattle in the region. This gap motivates this research.

\subsection{Research Objective}

%The objective of this research is to estimate disease prevalence in cattle and elk in the GYE by mechanistically modeling transmission between the species and incorporating how land-use change and landscape configuration influence disease transmission. This research develops a mathematical-epidemiological model, in combination with landscape ecology metrics, to estimate transmission rates between these species, with the goal of modeling disease spread between elk and cattle. Landscape ecology metrics are included into the mathematical-epidemiological model to determine how landscape configuration, defined as the shape and amount of habitat overlap, impact brucellosis transmission. Variation on the metrics parameters accounts for how land-use change translates to brucellosis prevalence in each species.

The objective of this research is to estimate disease prevalence in cattle and elk in the GYE by mechanistically modeling transmission between the species as a function of land-use change and landscape configuration. A mathematical-epidemiological model is developed that incorporates landscape ecology metrics to estimate transmission rates between elk and cattle. Landscape ecology metrics are included into the mathematical-epidemiological model to determine how landscape configuration, defined as the shape and amount of habitat overlap, impact brucellosis transmission. Variation of the parameters of these metrics allows us to investigate how land-use change translates to brucellosis prevalence in each species.

This chapter adds to the existing research by responding to the \citet{national2017revisiting} report, and incorporates its recommendations in a novel approach. The model developed is the first to describe brucellosis transmission between elk and cattle in the GYE. Estimations on the amount and shape of habitat overlap between the species are provided. The application of landscape ecology metrics into an epidemiological model for brucellosis is original. The study gives estimates for brucellosis prevalence levels in cattle and elk as a result of land-use change and landscape configuration in the region. The results provide insights into how land-use change impacts disease transmission; moreover, management strategies regarding spatially-temporally separating the species are drawn from these results, with various stakeholders' concerns taken into consideration.

\section{A Brucellosis Model for Cattle and Elk}
\label{sec:1}

The model to describe cross-species disease transmission between cattle and elk in the GYE is based on the following conclusions and recommendations made in the \citet{national2017revisiting} report.

\begin{itemize}
\item ``Conclusion 1: With elk now viewed as the primary source for new cases of brucellosis in cattle and domestic bison, the committee concludes that brucellosis control efforts in the GYA will need to sharply focus on approaches that reduce transmission from elk to cattle and domestic bison."

\item ``Recommendation 1: To address brucellosis in the GYA, federal and state agencies should prioritize efforts on preventing \textit{B. abortus} transmission by elk. Modeling should be used to characterize and quantify the risk of disease transmission and spread from and among elk, which requires an understanding of the spatial and temporal processes involved in the epidemiology of the disease and economic impacts across the GYA."

\item ``Recommendation 7: The research community should address the knowledge and data gaps that impede progress in managing or reducing risk of \textit{B. abortus} transmission to cattle and domestic bison from wildlife."

\item ``Recommendation 7A: Top priority should be placed on research to better understand brucellosis disease ecology and epidemiology in elk and bison, as such information would be vital in informing management decisions."
 
\item ``Recommendation 7C: Studies and assessments should be conducted to better understand the drivers of land use change and their effects on \textit{B. abortus} transmission risk." 

\end{itemize}

The model was developed using the $SIRS$ framework presented in the last chapter. Again, it is assumed that the disease-host populations are compartmentalized into different classes to denote different stages of infection. The model for cattle and elk is influenced by the model for bison and elk developed in \citet{dobson1996population}. However, the model is modified by changing the functional form of births in each species, excluding vertical transmission, and including landscape ecology metrics for estimating cross-species habitat overlap. Vertical transmission was initially included but because its inclusion complicated the mathematical analysis and did not provide any difference in qualitative behavior of the dynamics, it was omitted. The objective of this study was to determine how land-use change and landscape configuration impact disease prevalence, so using a simplified model, excluding vertical transmission, sufficed in this determination. 

In mathematical epidemiological modeling, estimating disease transmission rates allows for a better approximation of disease incidence. Although various methodologies are used to estimate disease transmission rates, this study incorporates landscape ecology metrics to quantify aspects of habitat overlap to determine cross-species disease transmission rates. The model is parametrically calibrated based on \citet{dobson1996population}, GIS analysis, and other literature \citep{national2017revisiting,xie2009disease,rayl2019modeling,rickbeil2019plasticity}.

\subsection{Model Description}

As in Chapter 2, let there be an area $(a)$ that exists completely inside an area $(z)$, and let the remainder of $(z)$ be the area $(b)$, so that $(z-a=b)$. Inside area ($a$), let there be an area $(o)$, called the contact zone or overlap, that contains all points within a fixed distance $(d)$, called the depth of the contact zone, to the edge of $(a)$ and $(b)$. Moreover, assume ($d$) is homogenous around the perimeter of ($a$). Variables and parameters subscripted with capital ``$C$" are reserved for the domestic cattle population, and variables and parameters subscripted with capital ``$E$" are reserved for the wild elk population. Consider that there exists a homogeneously distributed elk population $(N_E)$ that inhabits region $(a)$. Also, suppose that there is a cattle population $(N_C)$, homogeneously distributed, that is situated in region $(b+o)$. $(z)$ represents the Designated Surveillance Area (DSA) of brucellosis in the GYE specified by the NASEM. $(a)$ is the range over which elk are located in the surveillance area throughout a year, estimated from \citet{rickbeil2019plasticity} (see Figure (\ref{MIGRATION})). $(b+o)$ represents the range of all cattle herds in the DSA throughout a year, estimated from global information system (GIS) data about cattle distribution on private lands, USFS grazing allotments, and BLM grazing allotments in the DSA (see Figure (\ref{COMPOSITE})). The area of elk range that does not intersect with cattle distribution---and hence has no interaction---is called the core habitat ($c$), and the area that does overlap, where the mixing of the populations occurs, is denoted by $(o)$, so that $(a=o+c)$.

As this model captures the spread of brucellosis amongst cattle and elk, let the cattle population $(N_C)$ be compartmentalized into classes $(S_C)$ for susceptible cattle, $(I_C)$ for infectious cattle, and $(R_C)$ for recovered cattle, so ($N_C=S_C+I_C+R_C> 0$) and $(S_C\ge0$), ($I_C\ge0$) and ($R_C\ge0$) for all time ($t$). As well, let the elk population $(N_E)$ be divided into classes $(S_E)$ for susceptible elk, $(I_E)$ for infectious elk, and $(R_E)$ for recovered elk, so ($N_E=S_E+I_E+R_E> 0$) and $(S_E\ge0$), ($I_E\ge0$) and ($R_E\ge0$) for all time ($t$).

Assume that within-species transmission of brucellosis to elk occurs when a susceptible elk interacts with an infectious elk in the core habitat $(c)$ or on the overlap $(o)$, and there is successful transmission. Likewise, assume that within-species transmission of brucellosis to cattle occurs when a susceptible cow interacts with an infectious cow on habitat $(b)$ or on the overlap $(o)$, and there is successful transmission. Now, assume that cross-species transmission to a susceptible member of either species occurs when it encounters an infectious individual of the other species on the overlap region, and there is successful transmission. An initial level of brucellosis prevalence in the infectious class of each population is assumed. 

More extensively, let ($b_{EC}$) be the average number of cross-species contacts that an elk makes in unit time, and let ($f_{EC}$) be the probability that cross-species contact between an infectious elk and susceptible cow transmits infection. The transmission rate of the disease from elk to cattle is then the probability that a cross-species contact transmits infection times the number of contacts in unit time,
\[
\delta_C=b_{EC}f_{EC}N_E.
\]
A susceptible cow has ($\delta_C$) infective interactions with the elk population in unit time, of which a fraction ($I_E$) is with infectious elk. The number of new infective cattle caused by infection from an elk in unit time is
\[
\delta_{C}S_{C}\dfrac{I_E}{N_E}.
\]

Similarly, the transmission rate of the disease from cattle to elk is the probability that the cross-species contact transmits infection times the number of contacts in unit time,
\[
\delta_E=b_{CE}f_{CE}N_C.
\]
A susceptible elk has ($\delta_E$) infective interactions with the cattle population in unit time, of which a fraction ($I_C$) is with an infectious cow. The number of newly infected elk caused by infection from a cow in unit time is then
\[
\delta_{CE}S_E\dfrac{I_C}{N_C}.
\]
Within-species infections are modeled in a similar way. Let ($b_{CC}$) be the average number of contacts that cattle and elk make with their own species in unit time, and let ($f_{CC}$) and ($f_{EE}$)  be defined as the probabilities that contact of cattle and elk with their own kind transmits infection, respectively. The transmission rate is the probability that a contact transmits infection multiplied by the number of within-species contacts in unit time for cattle is then
\[
\beta_C=b_{CC}f_{CC}N_C.
\]
The within-species transmission rate for elk is defined in a similar way, 
\[
\beta_E=b_{EE}f_{EE}N_E.
\]

After a successful disease transmission, a formerly pathogen-free animal becomes infectious and is able to infect susceptible members of its own species at rate $(\beta_C)$ for cattle and $(\beta_E)$ for elk, and is able to cross-species infect at rate $(\delta_C)$ for cattle and $(\delta_E)$ for elk. Infectious cattle are assumed to recover at rate $(\gamma_C)$ and lose temporary immunity at rate $(\eta_C)$. Infectious elk are assumed to recover at rate $(\gamma_E)$ and lose temporary immunity at rate $(\eta_w)$. 

Both populations are assumed to grow logistically, with intrinsic growth rates of ($\alpha_C$) and ($\alpha_E$), and have carrying capacities, ($\kappa_C$) and ($\kappa_E$), be a proportion of habitat available for the cattle and elk populations, respectively. The cattle population is assumed to grow logistically since the majority of herds in the GYE are produced as open-range grazing calf-cow beef operations, which would be modulated by the area available for habitation.  Also, assume that individuals in all compartments reproduce susceptible offspring. Natural mortality occurs in all compartments at rate $(\sigma_C)$ for the cattle population and $(\sigma_E)$ for the elk population. Note that for this model, natural mortality of the cattle population includes ranchers harvesting cattle for slaughter or sale. Disease induced mortality by infection of brucellosis in these populations is considered insignificant \citet{treanor2007brucellosis} and is, therefore, not included in the model. Vertical transmission is also not considered. The brucellosis transmission model between cattle and elk is then:

\small
 \begin{eqnarray}\label{model1}
S_C' &=& \alpha_C N_C\left( 1-\dfrac{N_C}{k_C}\right) - \delta_C\left(\dfrac{o}{b+o}\right)S_C\left(\dfrac{o}{a}\right)\dfrac{I_E}{N_E} - \beta_C S_C\left(\dfrac{I_C}{N_C}\right) + \eta_C S_C - \sigma_C R_C \nonumber,\\
I_C' &=& \delta_C\left(\dfrac{o}{b+o}\right)S_C\left(\dfrac{o}{a}\right)\dfrac{I_E}{N_E}+\beta_C S_C\left(\dfrac{I_C}{N_C}\right) - (\sigma_C + \gamma_C) I_C \nonumber,\\
R_C' &=& \gamma_C I_C - (\eta_C + \sigma_C) R_C \nonumber,\\
S_E' &=&\alpha_E N_E\left( 1-\dfrac{N_E}{k_E}\right)  -\delta_E\left(\dfrac{o}{a}\right)S_E\left(\dfrac{o}{b+o}\right)\dfrac{I_C}{N_C} -\beta_E S_E\left(\dfrac{I_E}{N_E}\right)+ \eta_E S_E - \sigma_E R_E, \nonumber\\
I_E' &=& \delta_E\left(\dfrac{o}{a}\right)S_E\left(\dfrac{o}{b+o}\right)\dfrac{I_C}{N_C} + \beta_E S_E\left(\dfrac{I_E}{N_E}\right) - (\sigma_E + \gamma_E) I_E\nonumber,\\
R_E' &=& \gamma_E I_E - (\eta_E + \sigma_E) R_E, \\
& & S_C(0) = \text{ }N_C(0),\text{ } I_C(0)\text{ } =\text{ }0, \text{ }R_C(0) \text{ }=\text{ } 0, \text{ }N_C \text{ }=\text{ } S_C+I_C+R_C,\nonumber\\
& & S_E(0) = \text{ }N_E(0),\text{ } I_E(0)\text{ } =\text{ }0, \text{ }R_E(0) \text{ }=\text{ } 0, \text{ }N_E \text{ }=\text{ } S_E+I_E+R_E.\nonumber
\end{eqnarray}

\normalsize

Since the populations are assumed to be evenly distributed throughout their respective habitats, $(\frac{o}{a})$ is the proportion of ($S_E$) or ($I_E$) that is in ($o$) at any given time, and $(\frac{o}{b+o})$ is the proportion of ($S_C$) or ($I_C$) that is in ($o$) at any given time. The size of $(o)$ can be approximated with the empirically derived formulations of \[o = kd\mu\sqrt{a},\]   where  \[ \mu = \frac{\ell}{2\sqrt{\pi a}}\] is the shape index of the fragmented area, ($\ell$) is the length of the perimeter of the fragmented area, ($k=3.55$) is an empirically calculated scaling constant, and $(\pi=3.1415$) is the ratio of a circle's circumference to its diameter. By substituting the approximated size of the overlap area into the model and substituting ($b=z-a=z-c-o$), the model can be rewritten as

 \begin{eqnarray}\label{model2}
S_C' &=&  \alpha_C N_C\left( 1-\dfrac{N_C}{k_C}\right) - \delta_C\left[\dfrac{(kd\ell)^2}{4\pi a (z-c)}\right]S_C\dfrac{I_E}{N_E} - \beta_C S_C\left(\dfrac{I_C}{N_C}\right) - \sigma_C S_C + \eta_C R_C,\nonumber\\
I_C' &=& \delta_C\left[\dfrac{(kd\ell)^2}{4\pi a (z-c)}\right]S_C\dfrac{I_E}{N_E}+\beta_C S_C\left(\dfrac{I_C}{N_C}\right) - (\sigma_C + \gamma_C) I_C, \nonumber\\
R_C' &=& \gamma_C I_C - (\eta_C + \sigma_C) R_C, \nonumber\\
S_E' &=&\alpha_E N_E\left( 1-\dfrac{N_E}{k_E}\right)  -\delta_E\left[\dfrac{(kd\ell)^2}{4\pi a (z-c)}\right]S_E\dfrac{I_C}{N_C}-\beta_E S_E\left(\dfrac{I_E}{N_E}\right)- \eta_E S_E + \gamma_E R_E, \nonumber\\
I_E' &=& \delta_E\left[\dfrac{(kd\ell)^2}{4\pi a (z-c)}\right]S_E\dfrac{I_C}{N_C} + \beta_E S_E\left(\dfrac{I_E}{N_E}\right) - (\sigma_E + \gamma_E) I_E,\nonumber\\
R_E' &=& \gamma_E I_E - (\eta_E + \sigma_E) R_E,\\
& & S_C(0) = \text{ }N_C(0),\text{ } I_C(0)\text{ } =\text{ }0, \text{ }R_C(0) \text{ }=\text{ } 0, \text{ }N_C \text{ }=\text{ } S_C+I_C+R_C,\nonumber\\
& & S_E(0) = \text{ }N_E(0),\text{ } I_E(0)\text{ } =\text{ }0, \text{ }R_E(0) \text{ }=\text{ } 0, \text{ }N_E \text{ }=\text{ } S_E+I_E+R_E.\nonumber\\\nonumber
\end{eqnarray}

\normalsize

\subsection{Model Analysis}\label{sec1}

This section mathematically analyzes system (\ref{model2}), and an ecological interpretation of these results is provided in Section 3.3. When the equations in system (\ref{model2}) are added, the differential equations that represent the change of the total cattle and elk populations are
\[
N_C' = \alpha_C N_C \left(1-\dfrac{N_C}{k_C}\right)- \sigma_C N_C,
\]
\[
N_E' = \alpha_E N_E \left(1-\dfrac{N_E}{k_E}\right)- \sigma_E N_E,
\]
The solutions of the equations above are
\[N_C(t)=\dfrac{k_C(\alpha_C-\sigma_C)e^{\alpha_C c_1 k_C + \alpha_C t}}{e^{\sigma_C (c_1) k_C + \sigma_C t}-\alpha_C e^{\alpha_C (c_1) k_C + (\alpha_C) t}},\]
\[N_E(t)=\dfrac{k_E(\alpha_E-\sigma_E)e^{\alpha_E c_1 k_E + \alpha_E t}}{e^{\sigma_E (c_1) k_E + \sigma_E t}-\alpha_E e^{\alpha_E (c_1) k_E +( \alpha_E )t}},\]
for constant ($c_1$). If ($\alpha_i\le\sigma_i$), for ($i\in(C,E)$) then the ($i$) population is extinct. If the  ($\alpha_i>\sigma_i$), then the ($i$) population is a positive constant value. For the following analysis, assume the populations are constant.

%If the intrinsic birth rate is less than the death rate ($\alpha_i<\sigma_i$, for $i=C,E$) ,then the $i$ population is a negative constant and not biologically relevant. If  $\alpha_i=\sigma_i$, then the $i$ population is extinct.  If the  $\alpha_i>\sigma_i$, then the $i$ population is a positive constant value. For the following analysis, we will assume the populations are constant.

\subsubsection{Model Equilibria}
The following cases address the equilibria of system (\ref{model2}) in which subpopulations are equal to or greater than zero. Without loss of generality, the components the equilibria are in the order of $(S_C,I_C,R_C,S_E,I_E,R_E)$. 

\begin{itemize}

\item Case 1: Both subpopulations are equal to zero

\begin{itemize}
\item{}
If ($N_C=0$), and ($N_E=0$), then the only equilibria is
 \[
 Q_0=(0,0,0,0,0,0).
 \]
($Q_0$) always exists and is stable if ($\alpha_C\le\sigma_C$), ($\alpha_E\le\sigma_E$), and ($\alpha_B\le\sigma_B$).

\end{itemize}
\item Case 2: One subpopulations is equal to zero
\begin{itemize}
\item{}
If ($N_i=0$) and ($N_j>0$), for ($i,j$ $\in$ $(C,E)$), then system (\ref{model2}) reduces to the classical $SIRS$ model with a logistic growth functional form, and has all of its dynamics \citep{hethcote2000mathematics}.
\end{itemize}

\item Case 3: Both subpopulations are positive

\begin{itemize}

\item{}
If ($N_C>0$) and ($N_E>0$), then system (\ref{model2}) has the following disease-free equilibrium (DFE) (both species inhabit the GYE without brucellosis infection)
\[
Q_{1}=\left(\dfrac{k_C(\alpha_C-\sigma_C)}{\alpha_C},0,0,\dfrac{k_E(\alpha_E-\sigma_E)}{\alpha_E},0,0\right)
.\]
($Q_{1}$) exists if ($\alpha_C>\sigma_C$), ($\alpha_E>\sigma_E$), and is locally asymptotically stable if 
\begin{equation}
\frac{\beta_C+\beta_E+\sqrt{(\beta_C+\gamma_E+\sigma_E-\beta_E-\gamma_C-\sigma_C)^2+4{\Psi_C}{\Psi_E}}}{\gamma_C+\sigma_C+\gamma_E+\sigma_E}<1,
\label{inequality}
\end{equation}
for \[\Psi_i=\delta_i\left[\dfrac{(kd\ell)^2}{4\pi a (z-c)}\right]\] and $(i=C,E)$. As shown in Chapter 2, for (\ref{inequality}) to be satisfied, it is sufficient that 
\begin{equation}
|\gamma_C+\sigma_C| \ge |(\beta_C + \sqrt{{\Psi_C}{\Psi_E}})|.
\label{inequality2}
\end{equation}

However, if 
\begin{equation}
|\gamma_C+\sigma_C| < |(\beta_C + \sqrt{{\Psi_C}{\Psi_E}})|
\label{inequality3}
\end{equation}
then
\begin{equation}
|\gamma_C+\sigma_C+\gamma_E+\sigma_E| \ge |(\beta_C + \beta_E+\sqrt{{\Psi_C}{\Psi_E}})|
\label{inequality4}
\end{equation}
is necessary for (\ref{inequality}) to be satisfied.

\item{}

If ($N_C>0$), ($N_E>0$) and ($N_B>0$), then system (\ref{model2}) also has the following endemic (non-boundary) equilibrium (brucellosis is endemic to both populations), although the exact form is not known
\begin{equation}
Q_2=(S_C^*,I_C^*,R_C^*,S_E^*,I_E^*,R_E^*).
\label{endemic_equilibrium}
\end{equation}

The existence and stability arguments are demonstrated in Chapter 2. To summarize, the existence of $(Q_2)$ is shown by reducing system (\ref{model2}) and solving for the equilibria to yield

\begin{equation}\tilde{A}I_E^3+\tilde{B}I_E^2+\tilde{C}I_E+\tilde{D}=0,
\label{poly}
\end{equation}
where the coefficients of (\ref{poly}) are comprised of parameter values from system (\ref{model2}). The signs of the coefficients determine the number of positive real, negative real, and complex roots of (\ref{poly}).
The real and positive roots of (\ref{poly}) are biologically relevant and are equal to ($I_E^*$), which is a component of the biologically relevant (real and positive) equilibria of \eqref{model2}. In order to determine a biologically relevant non-trivial equilibrium, let
\begin{equation}
\beta_C=\gamma_C +\sigma_C
\label{assumptionC}
\end{equation}
and
\begin{equation}
\beta_E=\gamma_E +\sigma_E.
\label{assumptionE}
\end{equation}

Assuming (\ref{assumptionC}) and (\ref{assumptionE}), applying Descartes' rules of signs, and applying properties of the discriminant of cubic polynomials to (\ref{poly}) result in at least one real and positive root of (\ref{poly}).
Given (\ref{assumptionC}) and (\ref{assumptionE}), then there is at least one real and positive root of (\ref{poly}), and therefore one positive biologically relevant equilibrium of the system (\ref{model2}). Negative equilibria are not biologically relevant and do not satisfy the constraints of the variables. As shown in the next section, ($Q_2$) is stable if (\ref{assumptionC}) and (\ref{assumptionE}) are assumed.

\end{itemize}

\end{itemize}

\subsubsection{Basic Reproductive Number}

The basic reproduction number, the number of secondary disease cases caused by introducing a single infectious individual into a completely susceptible population, of system (\ref{model2}) is

\scriptsize
\begin{eqnarray}
\mathcal{R}_0=\frac{\beta_C(\gamma_E + \sigma_E) + \beta_E(\gamma_C + \sigma_C)+\sqrt{\left((\gamma_E + \sigma_E)\beta_C - (\gamma_C + \sigma_C)\beta_E\right)^2+4\Psi_E\Psi_C(\gamma_E + \sigma_E)(\gamma_C + \sigma_C)}}{2(\gamma_C+\sigma_C)(\gamma_E + \sigma_E)}.
\label{Rnot}
\end{eqnarray}
\\
\normalsize
If ($\mathcal{R}_0<1$) then the equilibrium $(Q_1)$, in Section 3.2.1, is stable. If ($\mathcal{R}_0>1$), then the equilibrium $(Q_1)$ is unstable. If ($\Psi_i = 0$), for ($i\in(E,C)$), then the expression for a single population
\begin{equation}
\mathcal{R}_0=\frac{\beta_C}{\gamma_C+\sigma_C}, 
\label{Rnotsingle}
\end{equation}
which can be greater to, equal, or less than one, given the magnitudes of the parameters. If ($\beta_C=\gamma_C+\sigma_C$) and ($\beta_E=\gamma_E+\sigma_E$), then there is at least one positive equilibria of the system, as negative equilibria are not biologically relevant and do not satisfy the constraints of the variables, and
the basic reproductive number reduces to
\begin{equation}
 \mathcal{R}_0=\frac{\beta_C\beta_E+\sqrt{\beta_C\beta_E\Psi_C \Psi_E}}{\beta_C\beta_E},
\label{Rnotdouble}
\end{equation}
which is always greater than 1; thus, the equilibrium ($Q_1$) is unstable, and the trajectories of system (\ref{model2}) tend towards at least one positive endemic equilibrium ($Q_2$).

\subsection{Simulations}

Analysis of the model allows identification of equilibria, and hence long term dynamics, of model (\ref{model2}).  Calibration of the model, followed by simulations,  allows investigation of disease dynamics along the convergence path. The following numerical solutions were conducted using ode15s in MATLAB R2019A. The simulations were conducted by setting parameter values to the estimates in Table (3.1), as certain parameters (depending on the numerical experiment) were varied while all other relevant parameter values were held constant.

%Although the analysis provides long term dynamics about model (\ref{model2}), calibrating the model with estimated parameter values allows for more current estimation of disease prevalence in the populations. The following numerical solutions were conducted using ode15s in MATLAB\_R2019A. The simulations were conducted by setting parameter values to the estimates in Table (\ref{BrucTable}), as certain parameters (depending on the numerical experiment) were varied while all other relevant parameter values were held constant.

The majority of the parameter values were estimated from literature, but the landscape parameters ($z$), ($a$), ($b$), ($\ell$), ($o$), ($d$), and ($\mu$) were determined using ArcGIS. The total landscape area $(z)$ for the simulations is the Designated Surveillance Area (DSA) of brucellosis in the GYE specified by the NASEM, the gray section in Figure (\ref{COMPOSITE1}). The habitat area of elk $(a)$ in the GYE is the amount of hectares over which elk are located in the surveillance area throughout a year, estimated from \citep{rickbeil2019plasticity}, see Figure (\ref{COMPOSITE1}). 

\begin{figure}[h!]
\centering
{\includegraphics[scale=.6]{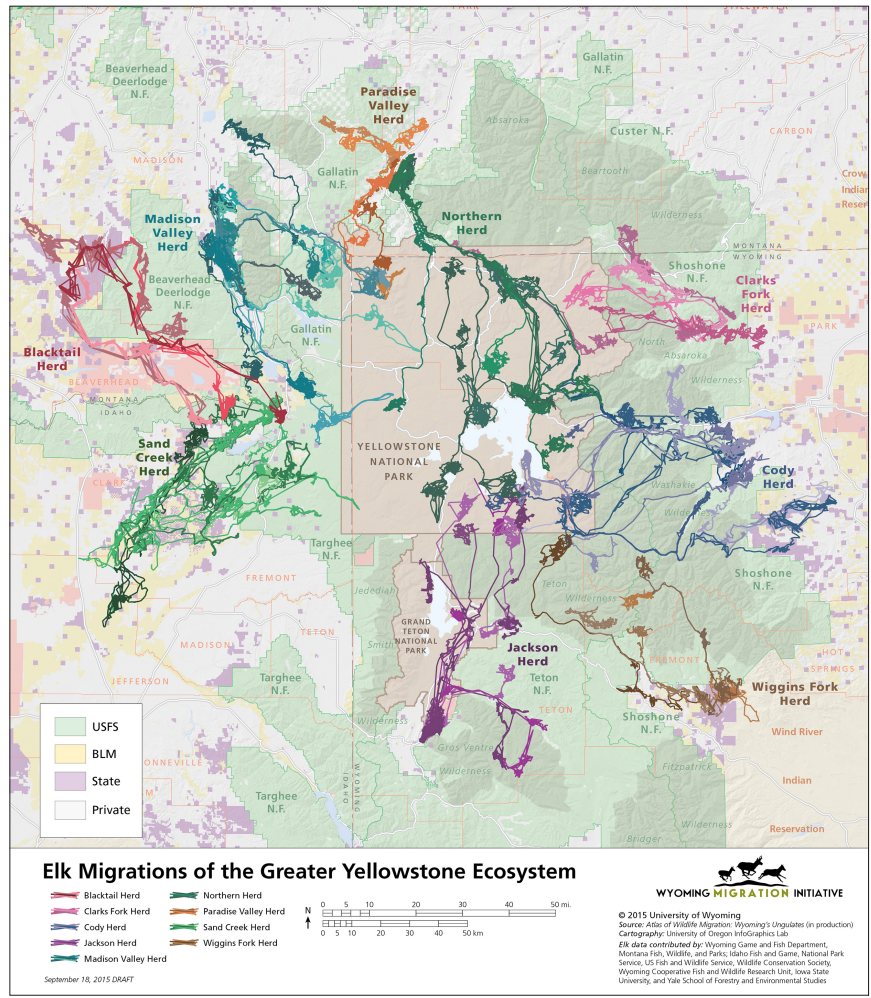}\label{MIGRATION1}}
\caption[Elk Migration Map in the GYE]{Map showing various elk herds migrations in the Greater Yellowstone Ecosystem. SOURCE: \citep{rickbeil2019plasticity}
}
\label{MIGRATION1}
\end{figure}

The value of ($a$) was determined by `georeferencing' Figure (\ref{MIGRATION1}) onto ArcGIS and drawing a perimeter line around the span of elk migration routes estimated from \citet{rickbeil2019plasticity}. Since the model assumes the populations are homogeneously distributed in their respective habitats, the perimeter of elk habitat was approximated by following the migratory boundaries of the elk herds in Figure (\ref{COMPOSITE1}). This resulted in the black outline that encompasses the entire Yellowstone National Park, the red area on (\ref{COMPOSITE}). Furthermore, Figure (\ref{COMPOSITE1}) shows how the estimated elk range fits into the DSA.

\begin{figure}[h!]
\centering
{\includegraphics[scale=.25]{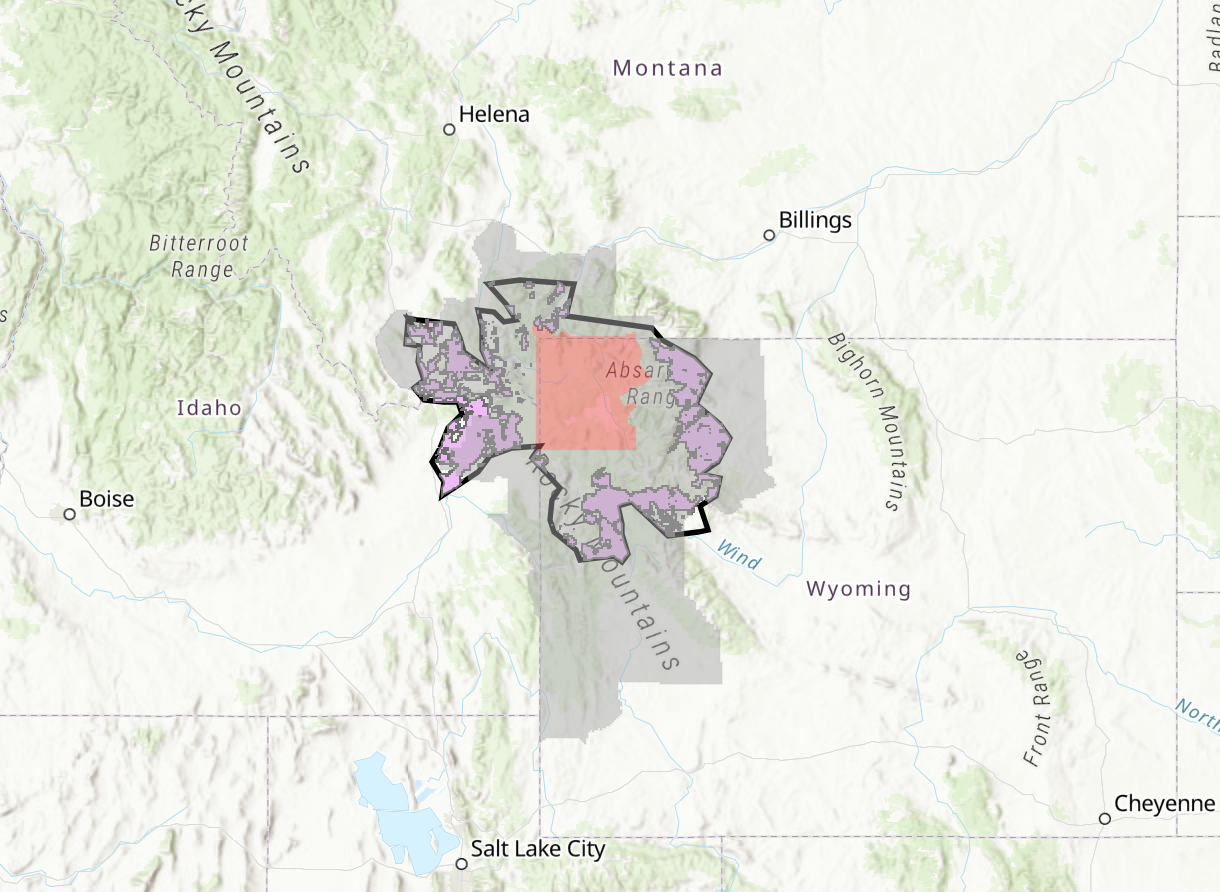}\label{COMPOSITE1}}
\caption[Cattle and Elk Distribution Map in the GYE]{Map showing the estimated elk range (the area within the black line), the designated surveillance area (the light gray area), cattle and elk habitat overlap area (the purple areas), and Yellowstone National Park (the red area).
}
\label{COMPOSITE1}
\end{figure}

The other landscape parameters can be derived from the landscape parameters ($z$), ($a$), and ($\ell$). The estimated range of cattle $(b=z - a)$. $(o)$ is the number of hectares that is mutual habitat for both species, the habitat overlap, the purple areas in Figure (\ref{COMPOSITE1}). To determine this value, BLM allotment data was overlaid with USFS allotment data, and private land allotment data onto $(z)$, and the intersection of those allotments with $(a)$ formed the total overlap area. ($\ell$) designates the perimeter length of the black outline in Figure (\ref{COMPOSITE1}), and ($\mu$) is the shape index of the black outline in Figure (\ref{COMPOSITE1}), calculated by the formula \[\mu = \frac{\ell}{2\sqrt{\pi a}}.\] The depth of the contact zone ($d$) was determined using the formula \[o = k d\mu \sqrt{a},\] which when solved for ($d$) yields \[d = \frac{o}{k \mu \sqrt{a}}.\] The core habitat of elk $(c)$ is the area within the black outline that does not contain the habitat overlap (the purple area).

\begin{landscape}

\begin{table}[h!] \begin{center}\scriptsize
\begin{tabular}{lllll} \toprule
\multicolumn{2}{l}{Parameter}              & Unit                      & Estimate        & Reference \\ \midrule
$S_i$ & Susceptible Host Population     & cattle/ elk                                 & 360,000/ 30,000 & \citep{national2017revisiting}        \\ 
$I_i$ & Infectious Host Population      & cattle/ elk                                 & 90,000/ 10,000 &   \citep{national2017revisiting}      \\
$R_i$ & Recovered Host Population       & cattle/ elk                                & 0/ 0 & Assumed       \\
$N_i$ & Host Population Size       & cattle/ elk                                &   450,000/ 40,000        & \citep{national2017revisiting}         \\

$\alpha_i$ & Host Birth Rate      & year$^{-1}$                                 & 0.142/ 0.07854 & \citep{rayl2019modeling, dobson1996population}     \\
$\kappa_i$ & Host Carrying Capacity          & cattle/ elk                                & 500,000/ 50,000 & \citep{xie2009disease, dobson1996population}  \\
$ \delta_i$ & Cross-Species Transmission Rate  & year$^{-1}$   &      0.1/ 0.1     &     \citep{xie2009disease, dobson1996population}         \\
$k$ & Scaling Constant             & unitless  &      3.55     &\citep{laurance1991predicting}       \\
$d$ & Depth of Contact Zone         & meter & 2.9583 &   \citep{rickbeil2019plasticity,USFSgrazing}$^{1}$   \\
$\ell$ & Perimeter Length           & meter  &     1,312,654.33       & \citep{rickbeil2019plasticity}    \\
$a$ & Elk Habitat Area      & hectare        &     9,784,751.05          & Derived from \citep{rickbeil2019plasticity}   \\
$z$ & Designated Surveillance Area     & hectare  &      15,629,689.60      &  \citep{national2017revisiting}      \\
$c$	  & Elk Core Habitat Area    & hectare         &  5,901,438.04                          & \citep{national2017revisiting}     \\ 
$ \beta_i $ & Within-Species Transmission Rate     &  year$^{-1}$   & 0.003/ 0.004     &       \citep{xie2009disease,dobson1996population}         \\
$\sigma_i$ & Host Mortality Rate         & year$^{-1}$   & 0.5/  0.15      &      \citep{xie2009disease,dobson1996population}          \\
$\gamma_i$ & Host Loss of Immunity Rate \!\!       & year$^{-1}$  & 0.5/ 0.5 &  \citep{,dobson1996population}   \\
$\eta_i$ & Host Recovery Rate \!\!       & year$^{-1}$  & 0.5/ 0.5 & \citep{xie2009disease,dobson1996population}  \\
$o$	  & Overlap Area (edge)     & hectare       &  3,883,313.01                          & \citep{rickbeil2019plasticity,USFSgrazing}$^{1}$\\ 
$b$	  & Cattle Habitat Area       & hectare &  5,844,938.55                         &\citep{USFSgrazing}$^{1}$         \\ 
$\mu$	  & Shape Index         & unitless    &  118.3778788                          & \citep{USFSgrazing}$^{1}$      \\ \bottomrule
\end{tabular}
\vspace{-.6cm}
\caption[Table of Parameters for Cattle-Elk Brucellosis Model]{Parameter definitions, units, and values for the model with $i\in(C,E)$. The values listed are the initial values used for simulations; some parameters varied through different simulations. $^{1}$The value was calculated by sourcing data from USFS and BLM into ArcGISPro.
\\}
\end{center} 
\label{BrucTable}
\end{table}

\end{landscape}

\normalsize

% the past 30 years, since brucellosis surveillance began, there has simultaneously been an increase in land-use change and a subsequent change of habitat overlap between elk and cattle. It is expected that in the next 40 years there will be an increase in land conversion and an expansion in the subsequent habitat overlap between the species, which translates to an increased shape index of the range that elk inhabit. It is assumed that when surveillance began, the habitat overlap shape index between cattle and elk was at the minimum value of one, and that there was a relatively linear increase in the shape index per year. Given this assumption, the shape index will reach 300 in the next 40 years.

Over the past 30 years, since brucellosis surveillance began, there has been substantial land-use change affecting habitat overlap between elk and cattle \citep{national2017revisiting}. It is expected that in the next 40 years land conversion will expand habitat overlap between the species, which translates to an increased shape index of the range that elk inhabit \citep{hansen2018trends}. It is assumed that when surveillance began, the habitat overlap shape index between cattle and elk was at the minimum value of one, and that there was a relatively linear increase in the shape index per year. Given this assumption, the shape index will reach 300 in the next 40 years.

%To facilitate the comparison of disease prevalence between each species, the size of the infected population of each species is presented as a proportion of that species' total population. The results are displayed with time-series plots to exemplify changes to the initial rate of infection, peak prevalence, and endemic prevalence, which are qualitative aspects of the time-series plots. The slope of a curve on a time-series plot from the initial time to the highest point on the curve indicates the initial rate of epidemic spread. The highest point of a curve on a time-series plot indicates the peak prevalence. Lastly, where a curve reaches equilibrium (flattens out) on a time-series plot indicates the endemic prevalence. 

To facilitate the comparison between each species, the size of the infected population of each species is presented as a proportion of that species' total population. The results are displayed with time series plots to illustrate changes to the initial rate of infection, peak prevalence, and endemic prevalence, which are qualitative aspects of the time series plots. The slope of a curve on a time series plot from the initial time to the highest point on the curve indicates the initial rate of epidemic spread. The highest point of a curve on a time series plot indicates the peak prevalence. Lastly, where a curve reaches equilibrium (flattens out) on a time series plot indicates the endemic prevalence.

Figure (\ref{CattleElkMu}a) and Figure (\ref{CattleElkMu}b) show the prevalence of brucellosis in the cattle ($I_C$) and elk ($I_E$) populations as the shape index of the habitat overlap changes. The shape index increases from ($\mu=1$), which corresponds to a circle, to ($\mu=300$), which corresponds to an irregular shape. The darkest blue line is the prevalence curve that corresponds to the shape index of ($\mu=1$), and the darkest red line is the prevalence curve that corresponds to the shape index of ($\mu=300$). The current estimated shape index ($\mu=118$) (approximately $\mu=120$) is a turquoise curve. As the shape index increases from ($\mu=1$) to ($\mu=300$), there is an increase in the initial rate of epidemic spread, peak prevalence, and endemic prevalence in both populations.

%%%%%%%%%%%%%%%%%%%%%%%%%%%%%%%%%%%%%%%%%%%%%%%%%%%%%%%%%%%%%%%%%%%%%%%%%%%%%%%%%%%%%%%%%%%%%%%%%%%%%%%%%%%%%%%%%%%%%%%%%%%%%%%%%%%%%%%%%%%%%%%%%%%%%%%%%%%%%%%%%%%%%%%%%%%%%%%%%%%%%%%%%%%%%%%%%%%%%%%%%%%%%%%%%%%%%%%%%%%%%%%%%%%%%%%%%%%%%%%%%%%%%%%%%%%%%%%%%%%%%%%%%%%%%%%%%%%%%%%%%%%%%%%%%%%%%%%%%%%%%%%%%%%%%%%%%%%%%%%%%%%%%%%%%%%%%%%%%%%

\newpage

\begin{figure}[!htbp]
\centering
%\hspace*{-.6cm}     
\subfloat[]{{\includegraphics[scale=.085]{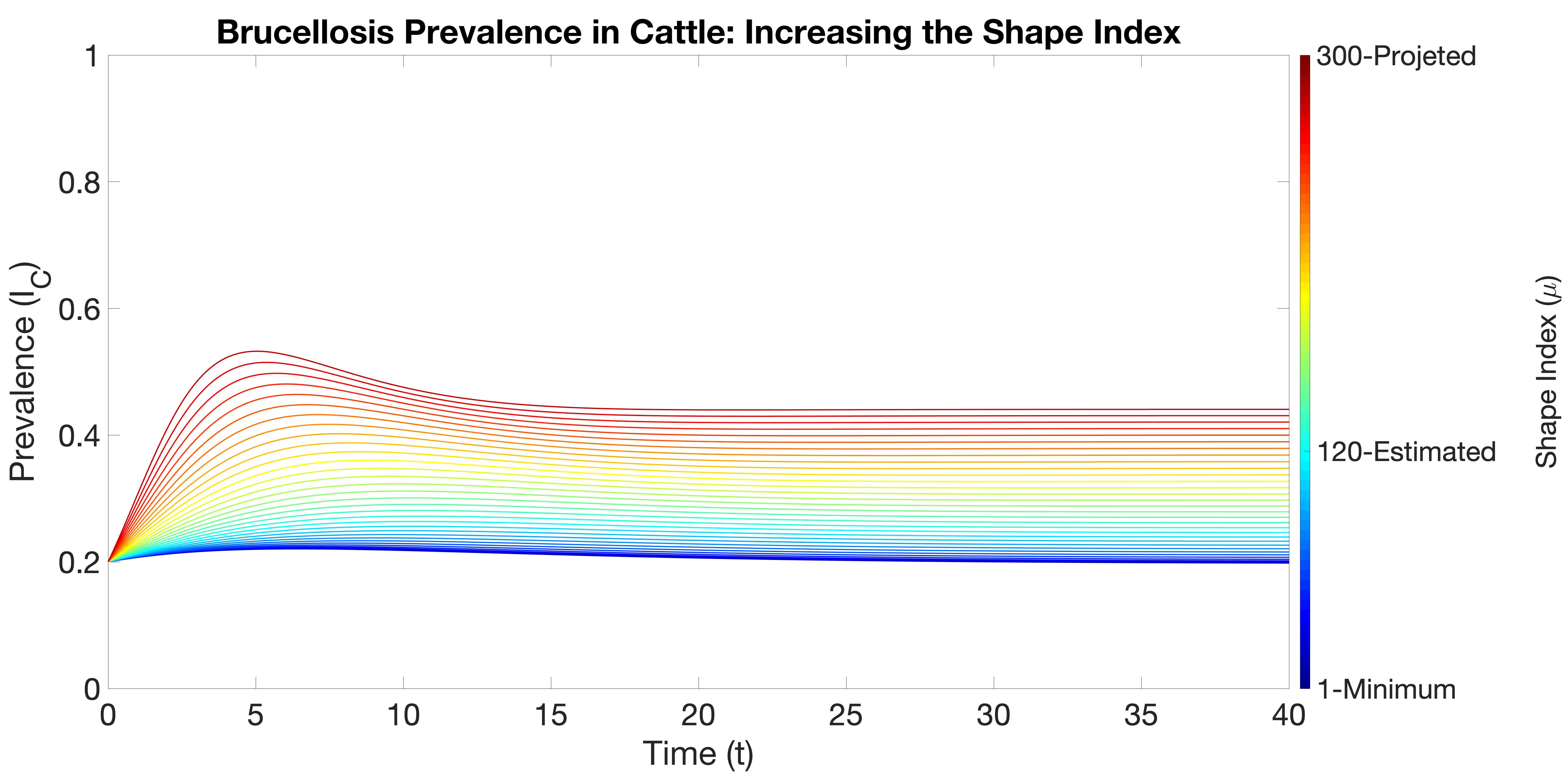}\label{fig:CattleChangeMu}}}

%\hspace*{-.6cm} 
%\subfloat[]{{\includegraphics[scale=.075]{images/ElkChangeMu}\label{fig:ElkChangeMu}}}
\caption%[time series plots of cattle and elk prevalence curves as change shape-index of contact zone]
[(a) Cattle Prevalence Curves as Change Shape Index of Contact Zone]{
(a) A time series plot showing the level of brucellosis prevalence in the cattle population as the shape index ($\mu$) increases from a minimum value of ($\mu=1$) (the darkest blue line) to an expected value assigned value of ($\mu=300$) (the darkest red line). 
%(b) A time series plot showing the level of brucellosis prevalence in the elk population as the shape index ($\mu$) increases from a minimum value of ($\mu=1$) (the darkest blue line) to an expected value assigned value of ($\mu=300$) (the darkest red line).
}
\label{CattleElkMu}
\end{figure}

\newpage

\begin{figure}[!htbp]
\ContinuedFloat
\centering
%\hspace*{-.6cm}     
%\subfloat[]{{\includegraphics[scale=.075]{images/CattleChangeMu}\label{fig:CattleChangeMu}}}

%\hspace*{-.6cm} 
\subfloat[]{{\includegraphics[scale=.085]{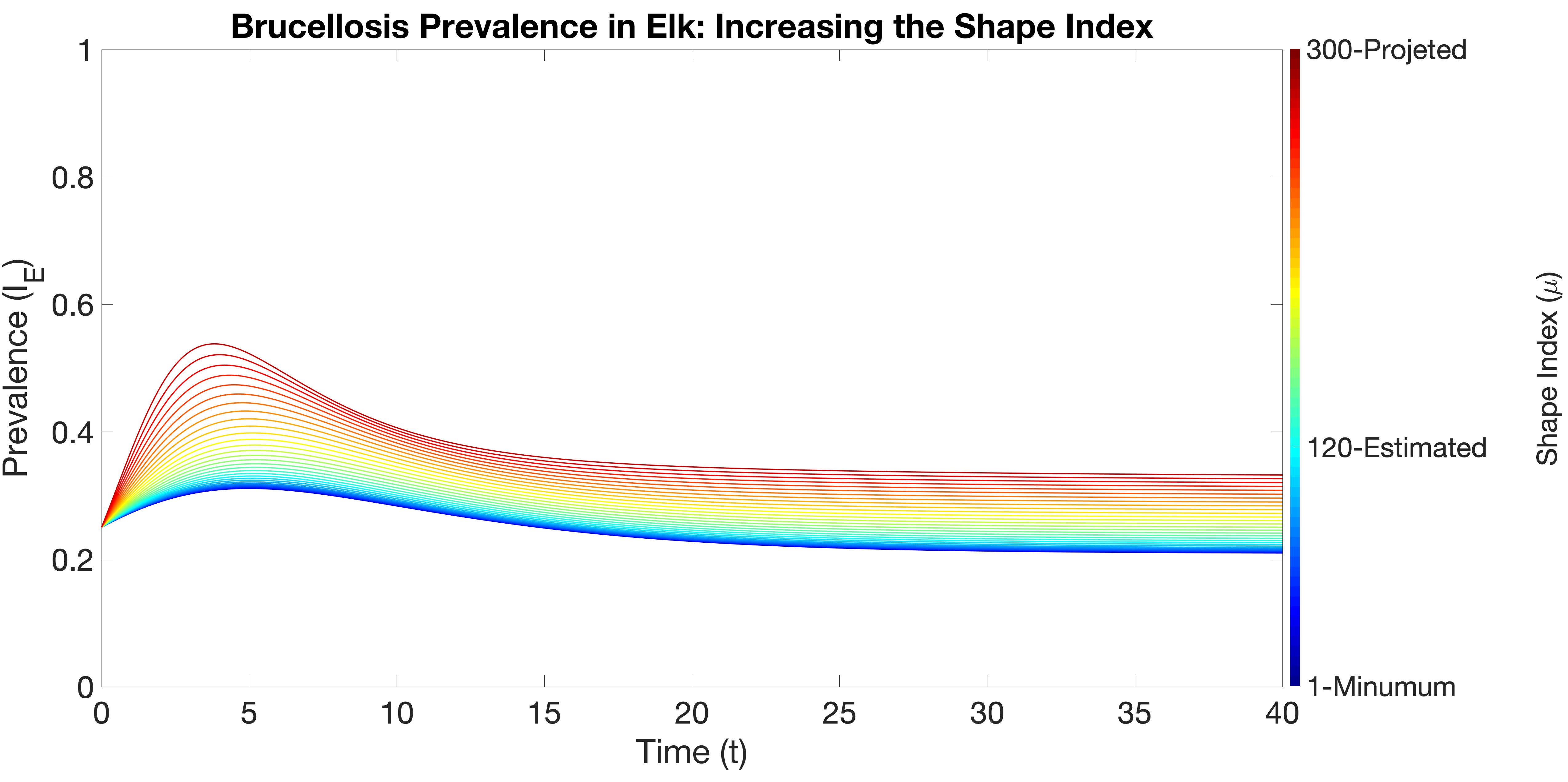}\label{fig:ElkChangeMu}}}
\caption%[time series plots of cattle and elk prevalence curves as change shape-index of contact zone]
[(b) Elk Prevalence Curves as Change Shape Index of Contact Zone]{
%(a) A time series plot showing the level of brucellosis prevalence in the cattle population as the shape index ($\mu$) increases from a minimum value of ($\mu=1$) (the darkest blue line) to an expected value assigned value of ($\mu=300$) (the darkest red line). 
(b) A time series plot showing the level of brucellosis prevalence in the elk population as the shape index ($\mu$) increases from a minimum value of ($\mu=1$) (the darkest blue line) to an expected value assigned value of ($\mu=300$) (the darkest red line).
}
\label{CattleElkMu}
\end{figure}

%%%%%%%%%%%%%%%%%%%%%%%%%%%%%%%%%%%%%%%%%%%%%%%%%%%%%%%%%%%%%%%%%%%%%%%%%%%%%%%%%%%%%%%%%%%%%%%%%%%%%%%%%%%%%%%%%%%%%%%%%%%%%%%%%%%%%%%%%%%%%%%%%%%%%%%%%%%%%%%%%%%%%%%%%%%%%%%%%%%%%%%%%%%%%%%%%%%%%%%%%%%%%%%%%%%%%%%%%%%%%%%%%%%%%%%%%%%%%%%%%%%%%%%%%%%%%%%%%%%%%%%%%%%%%%%%%%%%%%%%%%%%%%%%%%%%%%%%%%%%%%%%%%%%%%%%%%%%%%%%%%%%%%%%%%%%%%%%%%%

\begin{figure}[!htbp]
\centering
%\hspace*{-.6cm}     
\subfloat[]{{\includegraphics[scale=.085]{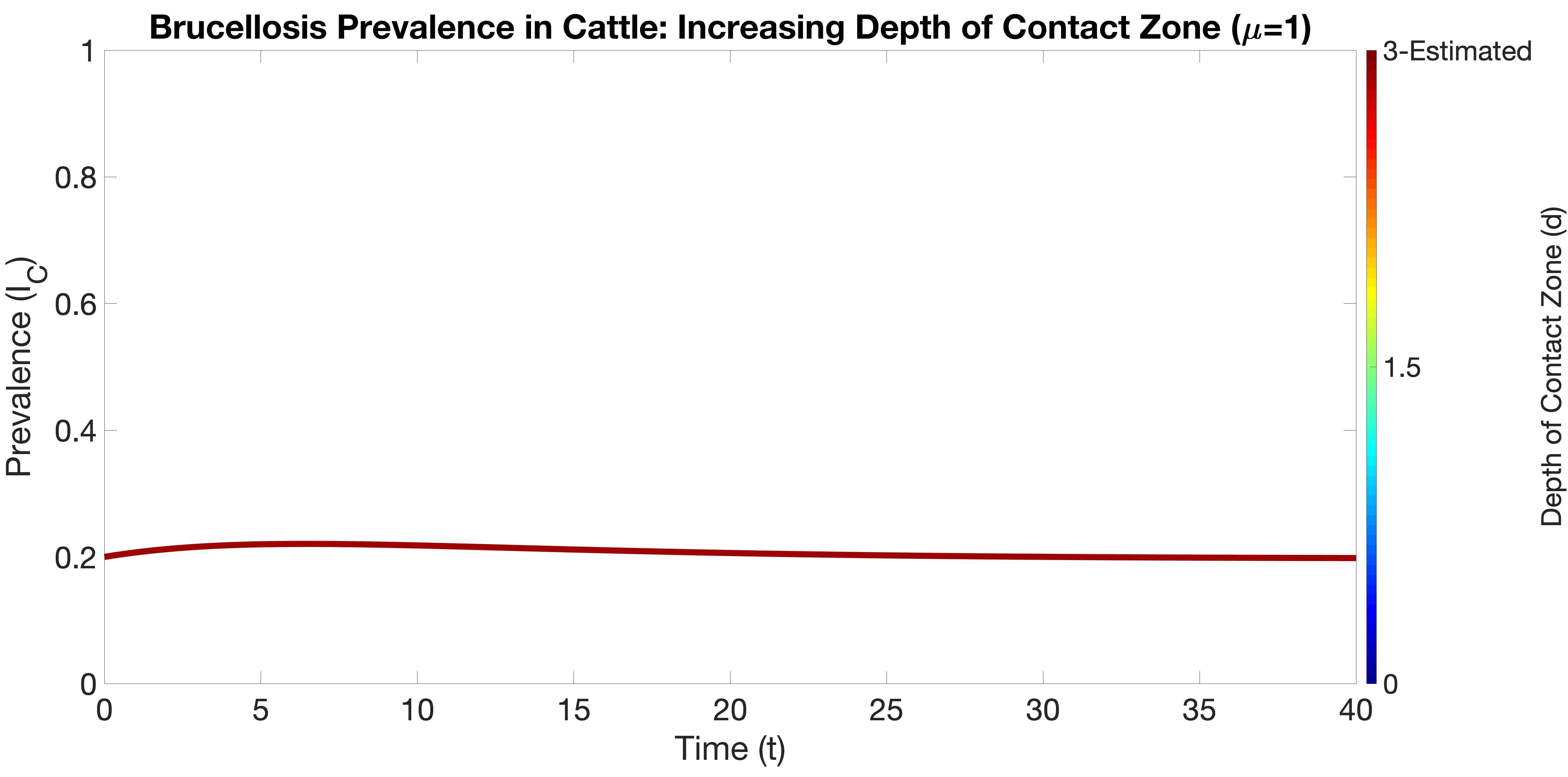}\label{fig:CattleChangeDMu1}}}

%\hspace*{-.6cm}     
%\subfloat[]{{\includegraphics[scale=.07]{images/ElkChangeDMu1}\label{fig:ElkChangeDMu1}   }}
\caption%[time series plots of cattle and elk prevalence curves as change depth of contact zone with low shape-index]
[(a) Cattle Prevalence Curves as Change Depth of Contact Zone with Low Shape Index]{
(a) A time series plot showing the level of brucellosis prevalence in the cattle population on a landscape where the mixing zone has the shape index of $(\mu=1)$ as the depth of the contact zone ($d$) increases from a minimum value of ($\mu=0$) to the estimated value of ($\mu=3$) (the darkest red line which overlays the other curves). 
%(b)  A time series plot showing the level of brucellosis prevalence in the elk population on a landscape where the mixing zone has the shape index of $(\mu=1)$ as the depth of the contact zone ($d$) increases from a minimum value of ($\mu=0$) to the estimated value of ($\mu=3$) (the darkest red line which overlays the other curves). 
}
\label{d_mu_1}
\end{figure}

For Figure (\ref{d_mu_1}), Figure (\ref{d_mu_118}), and Figure (\ref{d_mu_300}), three different cases of the shape index ($\mu=1$, $\mu=118$, and $\mu=300$) are analyzed. The depth of the contact zone ($d$) for each shape index varies from ($d=0$), indicated by the darkest blue curve on Figures (\ref{d_mu_1}), (\ref{d_mu_118}), and (\ref{d_mu_300}), to ($d=3$), which is the approximated average depth around the perimeter of the contact zone and corresponds to the darkest red curves on Figures (\ref{d_mu_1}), (\ref{d_mu_118}), and (\ref{d_mu_300}). % to the depth of the contact zone need to reach these levels of infection. These infection amounts were determined from the peak prevalence amounts in Figures (\ref{CattleElkMu}a,b) for the shape index of ($\mu=300$).

Figure (\ref{d_mu_1}a) and Figure (\ref{d_mu_1}b) show the prevalence of brucellosis in the cattle ($I_C$) and elk ($I_E$) populations as the depth of the contact zone ($d$) changes. The depth of the contact zone ranges from ($d=0$), corresponding to species not mixing, to ($d=3$).%, the depth that results in approximately 250,000 infections for cattle and 70,000 infections for elk.
The approximated average depth around the perimeter of the contact zone is ($d=3$), which is represented by a blue curve on Figures (\ref{d_mu_118}a) and (\ref{d_mu_118}b).% As the depth of the contact zone increases from ($d=0$) to ($d=1000$), there is an increase in the initial rate of epidemic spread, peak prevalence, and endemic prevalence in both the cattle and elk populations.

Figure (\ref{d_mu_118}a) and Figure (\ref{d_mu_118}b) show the prevalence of brucellosis in the cattle ($I_C$) and elk ($I_E$) populations as the depth of the contact zone ($d$) changes. The depth of the contact zone ranges from ($d=0$), corresponding to species not mixing, to ($\d=3$). The approximated average depth around the perimeter of the contact zone is ($d=3$), which is represented by a blue curve on Figure (\ref{d_mu_118}a) and Figure (\ref{d_mu_118}b). As the depth of the contact zone increases from ($d=0$) to ($d=3$), there is an increase in the initial rate of epidemic spread, peak prevalence, and endemic prevalence in both the cattle and elk populations.

Figure (\ref{d_mu_300}a) and Figure (\ref{d_mu_300}b) show the prevalence of brucellosis in the cattle ($I_C$) and elk ($I_E$) populations as the depth of the contact zone ($d$) changes. The depth of the contact zone ranges from ($d=0$), corresponding to species not mixing, to ($d=3$), which represents the approximated average depth around the perimeter of the contact zone. As the depth of the contact zone increases from ($d=0$) to ($d=3$), there is an increase in the initial rate of epidemic spread, peak prevalence, and endemic prevalence in both the cattle and elk populations.

\newpage

\newpage

\begin{figure}[!htbp]
\centering
%\hspace*{-.6cm}     
%\subfloat[]{{\includegraphics[scale=.07]{images/CattleChangeDMu118}\label{fig:CattleChangeDMu118}}}

%\hspace*{-.6cm}     
\subfloat[]{{\includegraphics[scale=.085]{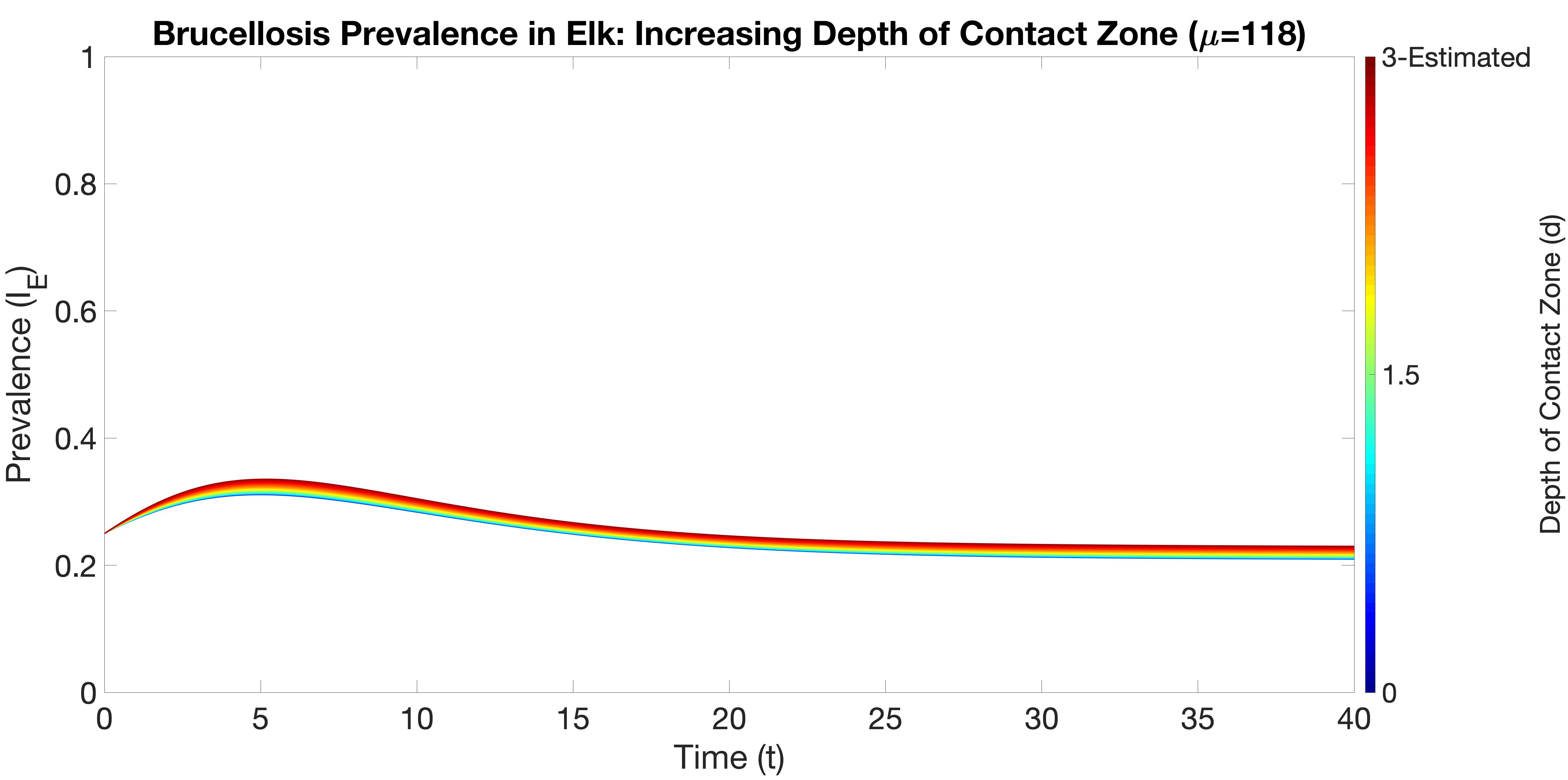}\label{fig:ElkChangeDMu118}   }}
\caption%[time series plots of cattle and elk prevalence curves as change depth of contact zone with medium shape-index]
[(b) Elk Prevalence Curves as Change Depth of Contact Zone with Medium Shape Index]{
%(a) A time series plot showing the level of brucellosis prevalence in the cattle population on a landscape where the mixing zone has the shape index of $(\mu=118)$ as the depth of the contact zone ($d$) increases from a minimum value of ($\mu=0$) (the darkest blue line) to the estimated value of ($\mu=3$) (the darkest red line). 
(b)  A time series plot showing the level of brucellosis prevalence in the elk population on a landscape where the mixing zone has the shape index of $(\mu=118)$ as the depth of the contact zone ($d$) increases from a minimum value of ($\mu=0$) (the darkest blue line) to the estimated value of ($\mu=3$) (the darkest red line). 
}
\label{d_mu_118}
\end{figure}

%%%%%%%%%%%%%%%%%%%%%%%%%%%%%%%%%%%%%%%%%%%%%%%%%%%%%%%%%%%%%%%%%%%%%%%%%%%%%%%%%%%%%%%%%%%%%%%%%%%%%%%%%%%%%%%%%%%%%%%%%%%%%%%%%%%%%%%%%%%%%%%%%%%%%%%%%%%%%%%%%%%%%%%%%%%%%%%%%%%%%%%%%%%%%%%%%%%%%%%%%%%%%%%%%%%%%%%%%%%%%%%%%%%%%%%%%%%%%%%%%%%%%%%%%%%%%%%%%%%%%%%%%%%%%%%%%%%%%%%%%%%%%%%%%%%%%%%%%%%%%%%%%%%%%%%%%%%%%%%%%%%%%%%%%%%%%%%%%%%

\newpage

\begin{figure}[!htbp]
\centering
%\hspace*{-.6cm}     
\subfloat[]{{\includegraphics[scale=.085]{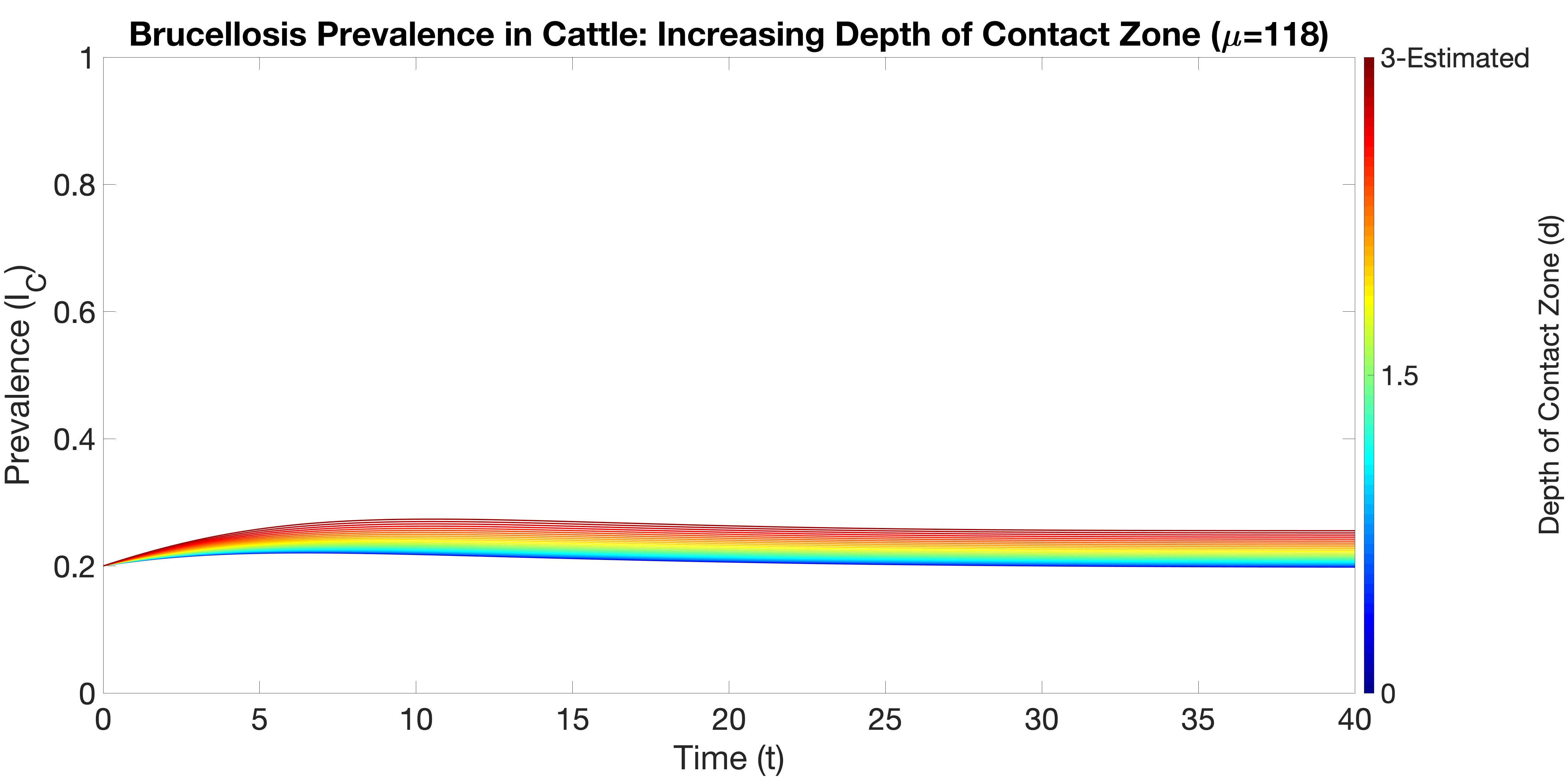}\label{fig:CattleChangeDMu118}}}

%\hspace*{-.6cm}     
%\subfloat[]{{\includegraphics[scale=.07]{images/ElkChangeDMu118}\label{fig:ElkChangeDMu118}   }}
\caption%[time series plots of cattle and elk prevalence curves as change depth of contact zone with medium shape-index]
[(a) Cattle Prevalence Curves as Change Depth of Contact Zone with Medium Shape Index]{
(a) A time series plot showing the level of brucellosis prevalence in the cattle population on a landscape where the mixing zone has the shape index of $(\mu=118)$ as the depth of the contact zone ($d$) increases from a minimum value of ($\mu=0$) (the darkest blue line) to the estimated value of ($\mu=3$) (the darkest red line). 
%(b)  A time series plot showing the level of brucellosis prevalence in the elk population on a landscape where the mixing zone has the shape index of $(\mu=118)$ as the depth of the contact zone ($d$) increases from a minimum value of ($\mu=0$) (the darkest blue line) to the estimated value of ($\mu=3$) (the darkest red line). 
}
\label{d_mu_118}
\end{figure}

\newpage

\begin{figure}[!htbp]
\ContinuedFloat
\centering
%\hspace*{-.6cm}     
%\subfloat[]{{\includegraphics[scale=.07]{images/CattleChangeDMu118}\label{fig:CattleChangeDMu118}}}

%\hspace*{-.6cm}     
\subfloat[]{{\includegraphics[scale=.085]{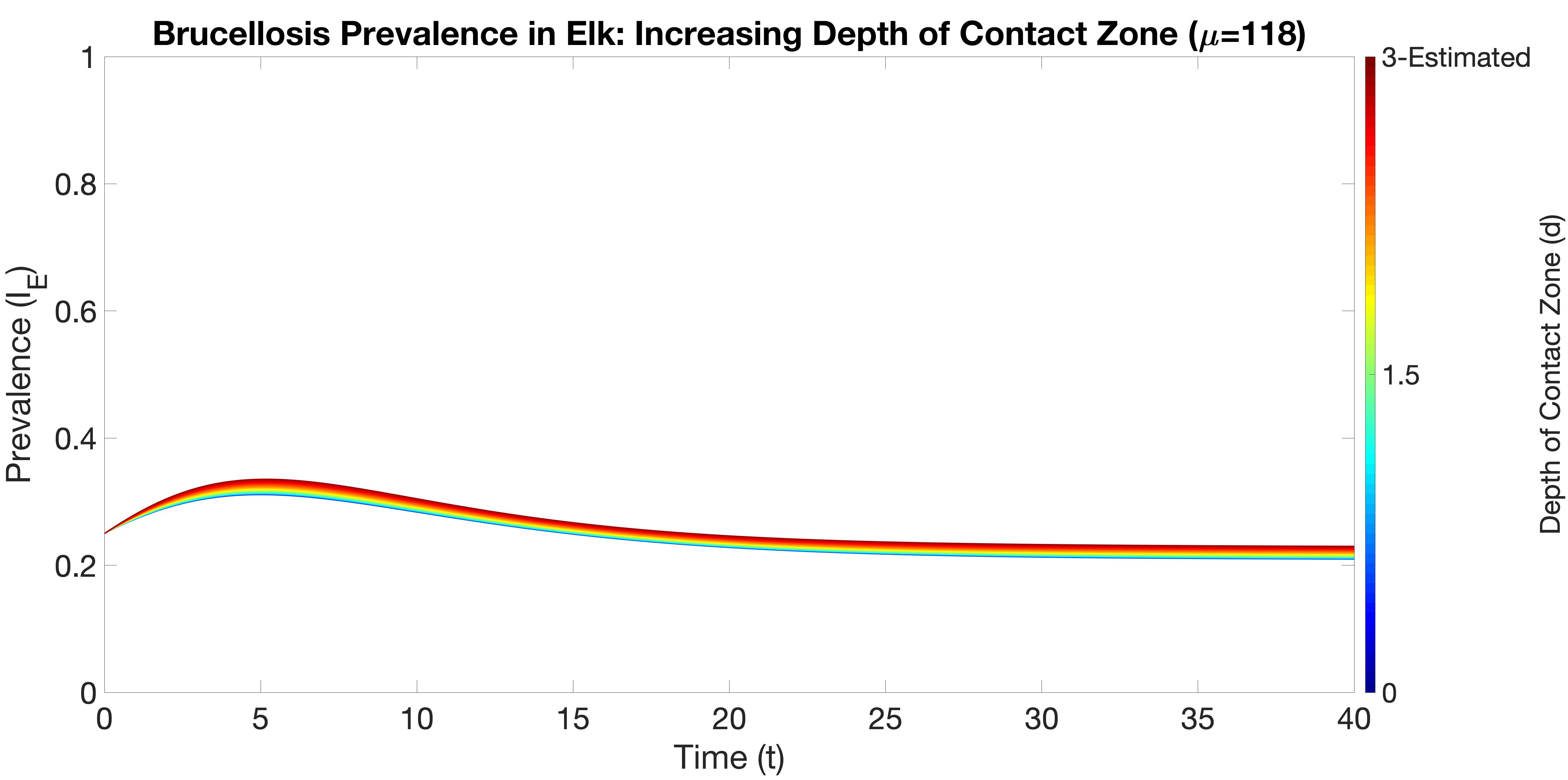}\label{fig:ElkChangeDMu118}   }}
\caption%[time series plots of cattle and elk prevalence curves as change depth of contact zone with medium shape-index]
[(b) Elk Prevalence Curves as Change Depth of Contact Zone with Medium Shape Index]{
%(a) A time series plot showing the level of brucellosis prevalence in the cattle population on a landscape where the mixing zone has the shape index of $(\mu=118)$ as the depth of the contact zone ($d$) increases from a minimum value of ($\mu=0$) (the darkest blue line) to the estimated value of ($\mu=3$) (the darkest red line). 
(b)  A time series plot showing the level of brucellosis prevalence in the elk population on a landscape where the mixing zone has the shape index of $(\mu=118)$ as the depth of the contact zone ($d$) increases from a minimum value of ($\mu=0$) (the darkest blue line) to the estimated value of ($\mu=3$) (the darkest red line). 
}
\label{d_mu_118}
\end{figure}

\newpage

\begin{figure}[!htbp]
\centering
%\hspace*{-.6cm}     
\subfloat[]{{\includegraphics[scale=.085]{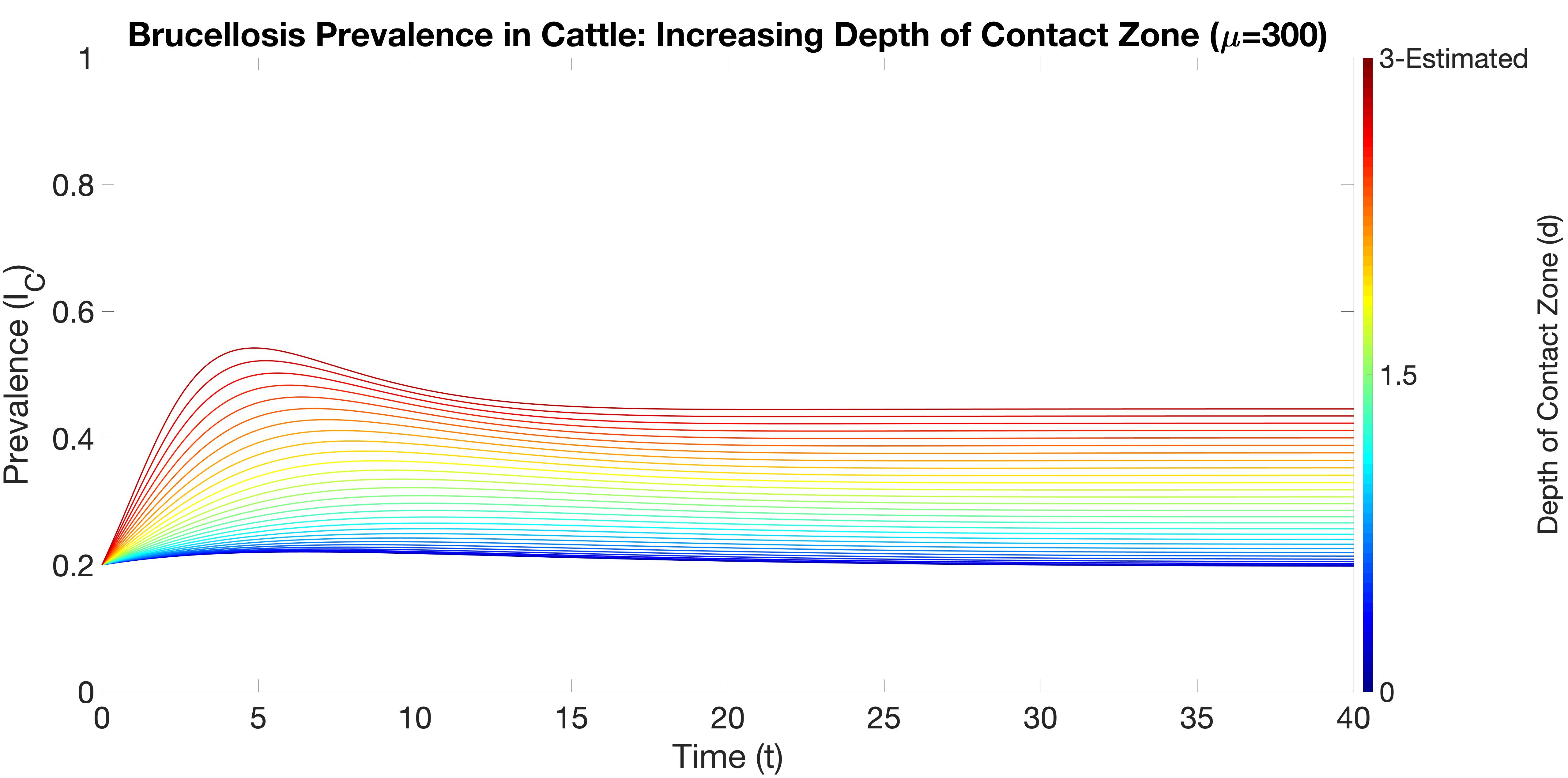}\label{fig:CattleChangeDMu300}}}

%\hspace*{-.6cm}     
%\subfloat[]{{\includegraphics[scale=.07]{images/ElkChangeDMu300}\label{fig:ElkChangeDMu300}   }}
\caption%[Time-series plots of cattle and elk prevalence curves as change depth of contact zone high shape-index]
[(a) Cattle Prevalence Curves as Change Depth of Contact Zone High Shape Index]{
(a) A time series plot showing the level of brucellosis prevalence in the cattle population on a landscape where the mixing zone has the shape index of $(\mu=300)$ as the depth of the contact zone ($d$) increases from a minimum value of ($\mu=0$) (the darkest blue line) to the estimated value of ($\mu=3$) (the darkest red line). 
%(b)  A time series plot showing the level of brucellosis prevalence in the elk population on a landscape where the mixing zone has the shape index of $(\mu=300)$ as the depth of the contact zone ($d$) increases from a minimum value of ($\mu=0$) (the darkest blue line) to the estimated value of ($\mu=3$) (the darkest red line). 
}
\label{d_mu_300}
\end{figure}

\newpage

\begin{figure}[!htbp]
\ContinuedFloat
\centering
%\hspace*{-.6cm}     
%\subfloat[]{{\includegraphics[scale=.07]{images/CattleChangeDMu300}\label{fig:CattleChangeDMu300}}}

%\hspace*{-.6cm}     
\subfloat[]{{\includegraphics[scale=.085]{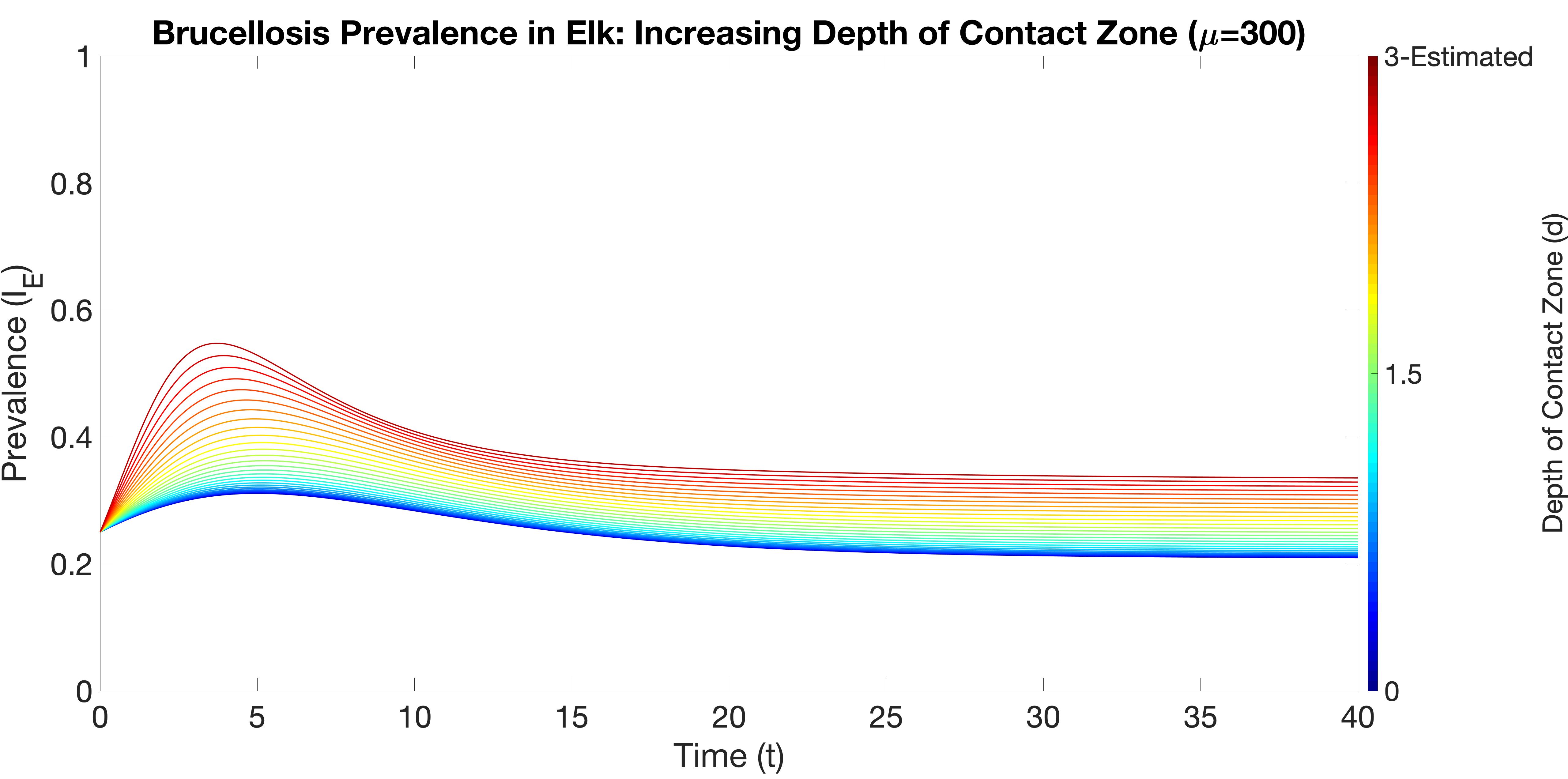}\label{fig:ElkChangeDMu300}   }}
\caption%[Time-series plots of cattle and elk prevalence curves as change depth of contact zone high shape-index]
[(b) Elk Prevalence Curves as Change Depth of Contact Zone High Shape Index]{
%(a) A time series plot showing the level of brucellosis prevalence in the cattle population on a landscape where the mixing zone has the shape index of $(\mu=300)$ as the depth of the contact zone ($d$) increases from a minimum value of ($\mu=0$) (the darkest blue line) to the estimated value of ($\mu=3$) (the darkest red line). 
(b)  A time series plot showing the level of brucellosis prevalence in the elk population on a landscape where the mixing zone has the shape index of $(\mu=300)$ as the depth of the contact zone ($d$) increases from a minimum value of ($\mu=0$) (the darkest blue line) to the estimated value of ($\mu=3$) (the darkest red line). 
}
\label{d_mu_300}
\end{figure}

%%%%%%%%%%%%%%%%%%%%%%%%%%%%%%%%%%%%%%%%%%%%%%%%%%%%%%%%%%%%%%%%%%%%%%%%%%%%%%%%%%%%%%%%%%%%%%%%%%%%%%%%%%%%%%%%%%%%%%%%%%%%%%%%%%%%%%%%%%%%%%%%%%%%%%%%%%%%%%%%%%%%%%%%%%%%%%%%%%%%%%%%%%%%%%%%%%%%%%%%%%%%%%%%%%%%%%%%%%%%%%%%%%%%%%%%%%%%%%%%%%%%%%%%%%%%%%%%%%%%%%%%%%%%%%%%%%%%%%%%%%%%%%%%%%%%%%%%%%%%%%%%%%%%%%%%%%%%%%%%%%%%%%%%%%%%%%%%%%%%%%%%%%%%%%%%%%%%%%%%%%%%%%%%%%%%%%%%%%%%%%%%%%%%%%%%%%%%%%%%%%%%%%%%%%%%%%%%%%%%%%%%%%%%%%%%%%%%%%%%%%%%%%%%%%%%%%%%%%%%%%%%%%%%%%%%%%%%%%%%%%%%%%%%%%%%%%%%%%%%%%%%%%%%%%%%%%%%%%%%%%%%%%%%%%%%%%%%%%%%%%%%%%%%%%%%%%%%%%%%%%%%%%%%%%%%%%%%%%%%%%%%%%%%%%%%%%%%%%%%%%%%%%%%%%%%%%%%%%%%%%%%%%%%%%%%%%%%%%%%%%%%%%%%%%%%%%%%%%%%%%%%%%%%%%%%%%%%%%%

\newpage

\section{Discussion}

The \citet{national2017revisiting} report recommended to model brucellosis transmission between elk and cattle, and to incorporate how land-use change and the subsequent landscape configuration contribute to brucellosis prevalence in the GYE. This study approached the problem by developing and analyzing a cross-species mathematical-epidemiological model and included landscape ecology metrics to estimate the cross-species transmission rates. To account for landscape configuration, the landscape ecology metrics used in this study are the depth and shape index of the contact zone between cattle and elk. Variation to the metric's parameters accounted for the amount of land-use change. The results indicate that increasing the depth and shape index of the contact zone between cattle and elk results in an increase of the initial rate of epidemic spread, peak prevalence, and endemic prevalence in both of the species. This is consistent with the results reported in \citet{cotterill2020disease}, but also extends those results to include the effect of habitat overlap between the species on brucellosis prevalence in cattle. In comparison to \citet{cotterill2020disease}, this study is a more generalized analysis of how land-use change, in a larger geographic range, affects brucellosis prevalence. The biological implications of the results are discussed in the following paragraphs.

As shown in Section 3.2.1, system (\ref{model2}) has multiple equilibria. The equilibrium ($Q_0$) represents the situation when both species go extinct. In the context of cattle production, this equilibrium is unrealistic; however, the existence of this equilibrium is an artifact of the model since the cattle population is assumed to grow logistically. Again, the assumption of logistic growth is made since the majority of cattle production in the GYE are open-range grazing operations, and elk is a free-roaming wild species. 
\newpage

($Q_1$) is the disease-free equilibrium, which is when both species inhabit the GYE without brucellosis infection. From conditions (\ref{inequality3}) and (\ref{inequality4}), if the cross-species ($\Psi_C$) and within-species ($\beta_C$) transmission coefficients for cattle are greater than their recovery ($\gamma_C$) and mortality ($\sigma_C$) rates, then, in order for brucellosis to become eliminated in both elk and cattle ($Q_1$), it is necessary that the recovery ($\gamma_i$)  and mortality ($\sigma_i$)  rates for cattle and elk be greater than or equal to the within-species ($\beta_i$)  and cross-species ($\Psi_i$) transmission coefficients of each species. However, if the recovery ($\gamma_C$) and mortality ($\sigma_C$) rates for only cattle are greater or equal to the within-species ($\beta_i$) and cross-species transmission coefficients for cattle, then this is sufficient to eliminate the disease in both species, as shown by condition (\ref{inequality2}). %The specific terms on the left-hand side and the right-hand side of conditions (\ref{inequality2}), (\ref{inequality3}), and (\ref{inequality4}) are a result of the sequence in-which the equations in system (\ref{model1}) have been ordered. If the elk population was written before the cattle population in system (\ref{model1}), then condition (\ref{inequality2}), for example, would indicate that in order to eliminate brucellosis in both elk and cattle populations, the rates of recovery and mortality associated with elk have to be greater than or equal to the within-species and cross-species transmission coefficients.

These results imply that for brucellosis to be controlled in the GYE, the degree to which the landscape parameters ($d$) and ($\mu$) are increased must be less than the rates at which cattle and elk are being removed from the population or recovering from infection. Additionally, if the degree to which the landscape parameters ($d$) and ($\mu$) are increased is less than the rates at which cattle are being removed from the population or recovering from infection, then both species will be driven to a disease-free state. If the depth of the habitat overlap is zero, then the sum of the removal rates must be greater than the same-species transmission coefficient for the disease-free equilibrium to be stable.

Even if cattle were to be vaccinated, and the within-species and cross-species transmission coefficients were reduced, the landscape parameters ($d$) and ($\mu$) inflate the cross-species transmission coefficient, which increases the likelihood of spill-over cases from elk to cattle. Given a likelihood of cross-species transmission, in order to drive the populations to a disease-free state, the amount and shape-index of the habitat overlap must decrease so that the rate of infection (cross-species combined with within-species) is less than the rates at which cattle and elk are removed from the population or recover from infection. Moreover, when the habitat overlap is more compactly configured, the populations of elk and cattle are more likely to co-exist without the threat of brucellosis infection.

If the landscape parameters increase to where neither condition (\ref{inequality2}) nor (\ref{inequality4}) are satisfied, then the disease-free equilibrium will be unstable (condition (\ref{inequality}) will not be satisfied). Additionally, if the sum of the removal rates equals the same-species transmission coefficient, ($\beta_C=\gamma_C+\sigma_C$) and ($\beta_E=\gamma_E+\sigma_E$), then neither condition (\ref{inequality2}) nor (\ref{inequality4}) can be satisfied, and the disease-free equilibrium ($Q_1$) will be unstable. Equations (\ref{assumptionC}) and (\ref{assumptionE}) represent that the total number of infections present in a species, due to within-species interactions, is equal to that species' recovery rate and mortality rate. 

Moreover, when equations (\ref{assumptionC}) and (\ref{assumptionE}) are assumed, there is always a level brucellosis prevalence in the cattle and elk populations. The level of infectivity may tend towards an equilibrium, ($Q_2$). Although the form of ($Q_2$) is not fully specified, ($Q_2$) is a theoretical endemic equilibrium, that can be shown to exist when equations (\ref{assumptionC}) and (\ref{assumptionE}) are assumed. When solving for an endemic equilibrium of system (\ref{model2}), equation (\ref{poly}) is formed. The signs of the coefficients of (\ref{poly}) determine the number of biologically relevant (real and positive) endemic equilibria, and depend on the relative magnitudes of the parameters of system (\ref{model2}).  As such, there is a high variability of the signs of the coefficients of equation (\ref{poly}); thus, any existence criteria about any biologically relevant endemic equilibria cannot be determined. However, if equations (\ref{assumptionC}) and (\ref{assumptionE}) are assumed, then the variability of the signs of the coefficients of (\ref{poly}) is reduced, and it can be shown that ($Q_2$) exists. The stability of ($Q_2$) can be determined by examining the basic reproductive number ($\mathcal{R}_0$) when equations (\ref{assumptionC}) and (\ref{assumptionE}) are assumed.

\newpage

The basic reproductive number ($\mathcal{R}_0$) is the number of secondary infections caused by a single infectious individual in a completely susceptible population. ($\mathcal{R}_0$) depends on, among other parameters, the shape index and the depth of the habitat overlap. As the landscape parameters increase, the basic reproductive number will increase to a value greater than one and the disease-free equilibrium will be unstable, meaning that disease prevalence will increase, and, as a result, the disease becomes endemic to the cattle and elk populations. If equations (\ref{assumptionC}) and (\ref{assumptionE}) are assumed, then the basic reproductive number will have the form of equation ($\ref{Rnotdouble}$), and $(\mathcal{R}_0)$ will be greater than one. Since ($Q_2$) exists under these assumptions, then ($Q_2$) can be asymptotically stable, as shown numerically (see Figure (\ref{CattleElkMu}), Figure (\ref{d_mu_1}), Figure (\ref{d_mu_118}), and Figure (\ref{d_mu_300})).

%Equations (9) and (10) can be assumed in order to examine the impact that the habitat overlap?s amount ($d$) and shape ($\mu$) have on cross-species disease transmission. If (9) and (10) are assumed, then brucellosis will become endemic to both populations. To minimize the impact of cross-species infection, then the landscape parameters ($d$ and $\mu$) would have to be minimized. For example, the shape of the habitat overlap would have to be as compact as possible. In essence, the landscape configuration of the habitat overlap would have to be as circular as possible and further land-use change would have to reduce the length of the edge between cattle and elk. Moreover, the depth of the habitat overlap would have to be minimized. This can be achieved either by building a fence to not allow the species to interact, or deterring elk from inhabiting cattle range. Even though the stability of an endemic equilibrium is not able to be mathematically deduced, system (2) can numerically be shown to have endemic levels of brucellosis prevalence. The following descriptions discuss how changes in the landscape parameters ($d$ and $\mu$) impact brucellosis prevalence in the cattle and elk populations.

Equations (\ref{assumptionC}) and (\ref{assumptionE})  can be used to examine the impact that the size ($d$) and shape ($\mu$) of habitat overlap have on cross-species disease transmission. If (\ref{assumptionC}) and (\ref{assumptionE})  are assumed, then brucellosis will become endemic to both populations. To minimize the impact of cross-species infection, then the landscape parameters ($d$) and ($\mu$) would have to be minimized. For example, the shape of the habitat overlap would have to be as compact as possible, and further land-use change would have to reduce the length of the edge between cattle and elk. Moreover, the depth of the habitat overlap would have to be minimized. This can be achieved either by building a fence to not allow the species to interact, or deterring elk from inhabiting cattle range. Even though the stability of an endemic equilibrium is not able to be mathematically deduced, system (\ref{model2}) can numerically be shown to have endemic levels of brucellosis prevalence. A description of the way that changes in the landscape parameters ($d$) and ($\mu$) impact brucellosis prevalence in the cattle and elk populations follows.

Figures (\ref{CattleElkMu}a) and (\ref{CattleElkMu}b) demonstrate the relationship between the shape of the habitat overlap ($\mu$) and brucellosis prevalence in the cattle and elk populations. As the shape index of the habitat overlap ($\mu$) is increased, as all other parameters are held constant, there is a simultaneous increase in the initial rate of epidemic spread, peak prevalence, and endemic prevalence in the cattle and elk populations. Moreover, as the shape index increases, the gap between the peak prevalence and the endemic prevalence increases. When the shape index of the habitat overlap ($\mu$) is less compact, more disease prevalence can be expected, resulting in a higher likelihood of spill over cases from elk to cattle in the region. Although it may not be the only factor driving disease prevalence, it can be seen that the shape of the habitat overlap ($\mu$) contributes to peak prevalence and endemic disease levels. The current value of ($\mu$) was determined by outlining elk foraging range (the span of their migration patterns), see Figure (\ref{COMPOSITE1}); thus, a change of $(\mu)$ would correspond to altered migration patterns for elk.

Land-use changes that shift the span of elk migration patterns change ($\mu$). Altered migration patterns of elk could result from factors that affect the availability of resources for elk, causing changes in elk foraging. These factors could be fencing, the instigation of new or more cattle production to the GYE, and other forms of land-use change in the region. Fencing may restrict elk migration and limit their access to resources, and increased cattle herd density may deplete resources, both of which cause the elk to forage in other locations. Moreover, an increase in the amount of winter feedgrounds for elk that are distributed in a spatially sparse manner could cause a shift in their congregations; consolidating where elk inhabit would adjust the shape of their foraging range as well as their population densities. Additionally, as more sections of the GYE are established for cattle production, this would introduce more cattle to the region and the landscape could become more habitable for elk, causing elk to forage in new places where interactions with cattle are more likely \citep{brennan2017shifting}. This would impact ($\mu$) more if the sectioning was especially sparsely \newpage \noindent distributed. These land-use changes, as discussed in Section 3.1.2, impact the spatial temporal separation of the cattle and elk.

As this study is concerned with how land-use change and the subsequent landscape configuration contributes to disease transmission, varying the parameter space of the shape index ($\mu$) allowed for the determination of the corresponding disease prevalence levels for each population. Since elk primarily forage in areas where cattle inhabit, an increase of land-use change, either through cattle ranching or the installment of winter feed grounds as specified in Section 3.1.2, could shift elk migration patterns and, as a result, alter ($\mu$) in such a way that there is more habitat overlap between the species, cross-species interaction, and a higher likelihood of cross-species brucellosis transmission. Hence, more land-use change in the region would translate to an increase of cross-species disease transmission, which reflects the findings of \citep{hansen2009species, wolfe2005bushmeat,cross2010probable}. Figure (\ref{CattleElkMu}) validates the discussion in Section 3.1.2.

Figure (\ref{d_mu_1}), Figure (\ref{d_mu_118}), and Figure (\ref{d_mu_300}) show the relationship between the depth of the habitat overlap and brucellosis prevalence in cattle and elk, respectively. In Figure (\ref{d_mu_1}), Figure (\ref{d_mu_118}), and (\ref{d_mu_300}), three landscape configurations ($\mu=1$), ($\mu=118$), and ($\mu=300$) are analyzed, and the depth of the contact zone ($d$) for each shape varies from ($d=0$) to ($d=3$), the current estimated depth. The shape index of the habitat overlap in Figure (\ref{d_mu_1}a) and Figure (\ref{d_mu_1}b) is ($\mu=1$). As the depth of the habitat overlap increases from ($d=0$) to ($d=3$), there is no change in the initial rate of epidemic spread, peak prevalence, and endemic prevalence for both the cattle and elk populations. Figure (\ref{d_mu_118}a) and Figure (\ref{d_mu_118}b) have a habitat overlap with the shape index of ($\mu=118$). As the depth of the habitat overlap increases from ($d=0$) to ($d=3$, there is a slight increase in the initial rate of epidemic spread, peak prevalence, and endemic prevalence for both the cattle and elk populations when compared to Figure (\ref{d_mu_1}a) and Figure (\ref{d_mu_1}b). This indicates that a more compact shape of the habitat overlap acts as a buffer from disease transmission. Figure (\ref{d_mu_300}) and Figure (\ref{d_mu_300}) assume a habit overlap with the shape index of ($\mu=300$). As the depth of the habitat overlap increases from ($d=0$) to ($d=3$), there is a higher increase in the in the initial rate of epidemic spread, peak prevalence, and endemic prevalence for both the cattle and elk populations when compared to a shape index of ($\mu=118$). The current value of ($d$) was determined from the amount of habitat overlap between elk and cattle in the foraging range of elk; thus, a change in ($d$) would correspond to a change in habitat overlap between the species.

Land-use changes that alter the amount of habitat overlap change ($d$). A change in the amount of the habitat overlap between cattle and elk could be caused by an introduction of wildlife fencing into new areas, and implementation of new feedgrounds for elk, or by an expansion or intensification of cattle ranching. Implementing wildlife fences would not allow elk to migrate onto areas where cattle inhabit, thus limiting cross-species interactions. The installment of more supplemental feedgrounds for elk does deter elk from foraging on cattle habitat which would decrease ($d$). However, the usage of the supplemental feedgrounds increases the relative abundance and density of elk, causing an increase in elk foraging which could potentially overlap with cattle habitat. A change in ($d$) may be caused by increasing the number of cattle in an area, thereby increasing in cattle herd density. This would increase the likelihood of cattle and elk interactions. Moreover, an expansion of cattle ranching operations could increase the habitat overlap between cattle and elk. Generally, cattle in this region are located on lowland grasslands, and the foraging preference of elk could increase the amount of habitat overlap, as both show a preference to the same type of habitat. As discussed in Section 3.1.2, the land-use changes specified above affect the spatial temporal separation of the cattle and elk.

Land-use change and its impacts on landscape configuration affect the spatial-temporal processes associated with the epidemiology of brucellosis and complicate spatial-temporal separation of the species, as discussed in Section 3.1.2. Since surveillance began in the GYE, there has been an increase in cattle ranching which has introduced cattle to more areas where elk forage, causing an increase in the amount of habitat overlap between the species, either by elk foraging for food or cattle being on or near elk migration routes. This translates to an increase in the depth of the habitat overlap, resulting in more cross-species interactions and brucellosis transmission. Land-use change and the subsequent landscape configuration in the GYE have increased the proximity of cattle and elk and, as such, there has been a sustained amount of spillover cases to cattle. Figure (\ref{d_mu_1}), Figure (\ref{d_mu_118}), and Figure (\ref{d_mu_300}) validates this claim and support the discussion in Section 3.1.2.

The significance of the results is that land-use change and the landscape configuration contribute to brucellosis being endemic to the both cattle and elk populations. Based on the \citet{national2017revisiting} recommendations to characterize the risk of brucellosis transmission from elk, and assess drivers of land-use change and their effects on disease spread, a model was constructed that had not been applied in this field before. The model builds upon the findings of \citet{rayl2019modeling} and provides a mechanistic explanation for how land-use change and the landscape configuration affect brucellosis transmission. The results also provide insight to related concerns in the GYE. As more cattle ranching is introduced, and elk population densities increase, either naturally or through artificial aggregation, there is an expected expanse of elk foraging range onto areas that are inhabited by cattle, thus amplifying the spread of brucellosis \citep{barasona2014spatiotemporal,brennan2017shifting}.

The results add to the current research by providing Brucellosis prevalence estimation in cattle and elk as a result of land-use change and the subsequent shape and amount of habitat overlap between the species in the GYE. \citet{dobson1996population} estimated brucellosis prevalence for bison and elk near Yellowstone National Park. \citet{cotterill2020disease} determined how the supplemental feedgrounds for elk impact brucellosis transmission in the elk populations and estimated disease prevalence at those locations. In comparison to \citet{cotterill2020disease}, this study is a more generalized analysis of how land-use with the addition of providing estimations for brucellosis prevalence in cattle, using data about the amount of habitat overlap between the species. The results provided in this dissertation consider a larger geographic region and factors that had not been examined by \citet{dobson1996population} and \citet{cotterill2020disease}. The results also demonstrate how epidemiological characteristics will change as a result of land-use change. This study explicitly shows how the shape and amount of the habitat overlap contributes to the initial rate of epidemic spread, peak prevalence, and endemic disease prevalence. The results are supported by \citep{patz2004unhealthy,kilpatrick2009wildlife,suzan2012habitat}.

\subsection{Management Strategies}

The shape of the contact zone between elk and cattle ($\mu$) can be a result of management actions taken. Changes to ($\mu$) can be attributed to altered migration patterns of elk. A possible land management strategy that could alter migration routes of elk may be the temporary suspension of the supplemental feedgrounds for elk. The suspension of these feedgrounds supports the findings of \citep{cotterill2020disease}. They found that supplementally feeding elk at these feedgrounds during the winter increases the brucellosis prevalence in their population and contributes to spillover cases in cattle. Moreover, although the feedgrounds were established to deter elk from interacting with cattle, their placement seems to be directing elk to migrate in particular directions and to congregate in particular regions; this results in a foraging range that has a non-compact shape and increases the amount of edge between the habitat of elk and cattle. The suspension of the feedgrounds could force elk to develop a more compact foraging range, at least until brucellosis is eradicated. Reducing ($\mu$) reduces the amount of habitat overlap between elk share and cattle.

In theory, reducing of the edge to area ratio of elk habitat would reduce the likelihood of cross-species infection. However, the simulations did not account for an increase in elk habitat area that may result from an expansion of foraging, as the numerical experiments assumed the area was held constant while the perimeter increased. Hence, as elk migrate during the winter season to lowland grasslands where they can interact with cattle, if the supplemental feedgounds are removed and the elk no longer congregate in particular regions, elk would be expected to have more interactions with cattle. Thus, in order to diminish the likelihood of cross-species transmission, cattle and elk may also need to be separated at the lowland grasslands in the winter season.

In conjunction with reducing ($\mu$), the model indicates that brucellosis prevalence throughout the ecosystem might also be achieved by reducing ($d$) through land management. By restricting cattle movements through, for example, the use of feedlots during the season that brucellosis spillovers are most likely, there can be a reduction in disease spillover. Another possible solution may be to build wildlife fence or a buffer zone between private land utilized for cattle ranching and public lands where elk forage. Cattle would continue to be confined inside the converted lands, while elk would be restricted from entering and migrating throughout. The construction of a veterinary cordon fence between elk and cattle would minimize ($d$). Although this would be a very effective method at mitigating transmission between the species, there are potential negative consequences of its employment. The establishment of wildlife fencing would be very expensive to install and the maintenance could potentially cost even more. Moreover, creating a buffer zone or wildlife fence would disrupt to the migration patterns of elk, potentially negatively affecting the ecosystem. These control efforts exhibited to all species may contribute to eliminating the disease from the GYE, not only from cattle but also in wildlife.

\section{Conclusion}

Brucellosis incidence in the Greater Yellowstone Ecosystem continues to be a persistent problem for cattle and elk populations, and is of concern to both cattle-ranchers and wildlife-management entities. The \citet{national2017revisiting} report concluded that understanding the spatial and temporal processes associated with brucellosis in the GYE and incorporating landscape ecology models could contribute to addressing the epidemiology of brucellosis. To address these matters, a differential equations compartmental model was developed to describe brucellosis transmission between cattle and elk populations in the GYE as a function of landscape ecology metrics. This study used landscape metrics in congruence with GIS data about cattle distribution to understand how land-use across a spatially heterogeneous ecosystem plays a role in disease transmission. Few studies to date have neither mechanistically modeled brucellosis transmission between elk and cattle in the GYE, nor incorporated how landscape configuration, by way of habitat overlap amount and shape, has contributed to disease prevalence.

This chapter adds to the existing research by presenting the first model to describe brucellosis transmission between elk and cattle in the GYE as a result of land-use change. It uniquely uses aspects of land-use change and landscape configuration to determine brucellosis prevalence in the GYE. This model was developed to address recommendations posed by \citet{national2017revisiting} in novel approach, and serves as an extension and application of the model developed in the previous chapter. This model was parameterized for specific populations and a particular disease. The study gives estimates for brucellosis prevalence levels in cattle and elk as a result of land-use change and landscape configuration in the region. The results provide insights into how land-use change impacts disease transmission the application of the framework demonstrates its ability to approximate disease prevalence as a result of land-use change, as well as furthers the understanding about the relationship between patterns of landscape configuration and disease spread. Furthermore, implications for brucellosis management that were drawn from the results of the model were discussed, with various stakeholders' interests considered.

This work also contributes to the literature of ecology and epidemiology. It provides an example of how landscape ecology metrics can be integrated with an infectious disease model, and how demonstrates a connection between landscape configuration and disease spread. As spatial-temporal separation of individuals is essential in mitigating the transmission of infectious diseases, the study provides insight as to how that can be more adequately achieved. Although, there are other aspects of disease spread and brucellosis persistence in the GYE which this study does not consider, it can be extended to examine how the inhabitance of bison factor into spillover cases consistently infecting cattle to the region. An extension of the model can further address how land-use change and landscape configuration systematically propagates disease spread throughout different patches in an ecosystem. This chapter is a contextual first step towards that understanding.

\bibliographystyle{spbasic}      % basic style, author-year citations

\bibliography{citations.bib}   % name your BibTeX data base

\end{document}